\documentclass[aps,prc,preprint,showpacs,showkeys]{revtex4}
\usepackage{amssymb}
\usepackage{amsmath}
\usepackage{graphicx}
\usepackage{bm}
\usepackage{epsf}

\begin{document}

\title{Effects of the symmetry energy on properties of neutron star crusts
near the neutron drip density}
\author{S. S. Bao}
\affiliation{School of Physics, Nankai University, Tianjin 300071, China}
\author{J. N. Hu}
\affiliation{School of Physics, Nankai University, Tianjin 300071, China}
\author{Z. W. Zhang}
\affiliation{School of Physics, Nankai University, Tianjin 300071, China}
\author{H. Shen}~\email{shennankai@gmail.com}
\affiliation{School of Physics, Nankai University, Tianjin 300071, China}

\begin{abstract}
We study the effects of the symmetry energy on the neutron drip density
and properties of nuclei in neutron star crusts. The nonuniform matter
around the neutron drip point is calculated by using the Thomas--Fermi
approximation with the relativistic mean-field model.
The neutron drip density and the composition of the crust are
found to be correlated with the symmetry energy and its slope.
We compare the self-consistent Thomas--Fermi approximation
with other treatments of surface and Coulomb energies,
and find that these finite-size effects play an essential role
in determining the equilibrium state at low density.
\end{abstract}

\maketitle


\section{Introduction}
\label{sec:1}

Neutron star crusts are important laboratories for the study of asymmetric
nuclear matter at subnuclear density~\cite{Cham08,PR00,PR07}.
The crust is divided into an outer crust and an inner crust at the neutron drip
density $n_{\rm{drip}} \sim 4 \times 10^{11}\,\rm{g\,cm^{-3}}$
where neutrons begin to drip out of nuclei~\cite{Stei98}.
It is well known that the outer crust consists of a lattice of nuclei
with a gas of electrons, while the inner crust contains neutron-rich nuclei,
dripped neutrons, and relativistic electrons~\cite{Cham08,PR00,PR07}.
Great efforts have been devoted to the study of neutron star crusts
because of their importance in astrophysical observations and complex phase
structure~\cite{Cham08,PR00,PR07,Stei98,Haen00,Haen01,Pote13,Maru13}.
At low densities around $n_{\rm{drip}}$,
the stable shape of the nucleus is spherical, but it may change from droplet
to rod, slab, tube, and bubble, known as nuclear pasta phases,
at relatively high densities~\cite{Rave83,Wata00,Oyam07,Bao14}.
The neutron drip density is determined by the neutron chemical
potential, which is strongly dependent on the nuclear symmetry energy
and its density dependence.
Therefore, it is interesting to investigate the effects of the symmetry
energy on the neutron drip density and properties of neutron star crusts
around the neutron drip point.

The nuclear symmetry energy and its density dependence play a crucial
role in understanding various phenomena in nuclear physics and
astrophysics~\cite{PR07,LiBA08}.
The symmetry energy $E_{\rm sym}$ at saturation density
can be constrained by experiments to be around $30\pm 4$ MeV,
while the symmetry energy slope $L$ at saturation density
is still very uncertain and may vary from about $20$
to $115$ MeV~\cite{Chen13}. Many properties of neutron stars,
such as the crust structure, the crust-core transition,
and the star radius, are sensitive to the symmetry energy
and its density dependence~\cite{Oyam07,Duco11,Mene11}.
In Ref.~\cite{Oyam07}, the properties of nuclei in the inner
crust were studied using a parametrized Thomas--Fermi approach;
they were found to be sensitive to the density dependence of
the symmetry energy. In Ref.~\cite{Gril12}, a self-consistent Thomas--Fermi
approximation was used to calculate the properties of the inner
crust including pasta phases, and it was found that $L$ could
have dramatic effects on the crust structure.

The equilibrium state of neutron star crusts can be determined by
minimizing the total energy density at a given average baryon density $n_b$
under the conditions of $\beta$ equilibrium and charge neutrality.
The outer crust is well described based on experimental
masses of neutron-rich nuclei, but the inner crust has to be
studied by using phenomenological models due to the presence of
dripped neutrons. In past decades, the structure of the inner crust
has been investigated by using various methods, such as the liquid-drop
model~\cite{Wata00,BBP71,RBP72} and the Thomas--Fermi
approach~\cite{Oyam07,Gril12,Mene08,Mene10}.
Using the Wigner--Seitz approximation, the crust is divided into
spherical cells, in which a nucleus is located in
the center surrounded by a gas of electrons and neutrons.
A simple treatment for the matter inside the Wigner--Seitz
cell is referred to as the coexisting phases (CP) method~\cite{Bao14,Mene08},
in which the matter inside the cell separates into a dense phase
and a dilute phase with a sharp interface. The two coexisting phases
satisfy Gibbs conditions for phase equilibrium, which correspond
to bulk equilibrium without finite-size effects.
The surface and Coulomb energies are perturbatively taken into
account after the coexisting phases are achieved.
Another treatment of the inner crust is based on a compressible
liquid-drop (CLD) model and in this treatment the equilibrium state
is determined by minimization of the total energy density including
the surface and Coulomb energies~\cite{BBP71,Latt85,Latt91}.
Therefore, the finite-size effects due to the surface and Coulomb
energies are properly taken into account in this method.
The Thomas--Fermi (TF) approximation is considered to be self-consistent
in the treatment of finite-size effects and nucleon distributions
and has been widely used in atomic and nuclear physics~\cite{TF07}.
The TF approximation has been used to study neutron star
crusts including pasta phases at zero temperature~\cite{Gril12,Mene09}
and finite temperature~\cite{Mene10}.
It is important to compare and analyze the differences between
these methods and explore their validity at low density.

This paper has two aims. The first one is to analyze the differences
between the methods used for the study of neutron star crusts,
so as to examine the finite-size effects due to the surface
and Coulomb energies in determining the equilibrium
state at low density. In fact, we find that
the energy density obtained in the CP method is generally
larger than that of the corresponding homogeneous phase at low densities
around the neutron drip point. The failure of
the CP method at low densities may be due to the improper
treatment of the surface and Coulomb energies.
The second aim of this paper is to investigate the effects of the symmetry
energy on the neutron drip density and properties of neutron
star crusts. To calculate the properties of neutron star crusts,
we employ the TF approximation, which is considered to be self-consistent
in the treatment of finite-size effects and nucleon distributions.
For the nuclear interaction,
we adopt the relativistic mean-field (RMF) theory,
which has been successfully used to study various phenomena
in nuclear physics~\cite{Sero86,Ring90,Meng06}.
In the RMF approach, nucleons interact via the exchange of
scalar and vector mesons, while the parameters
are fitted to nuclear matter saturation properties or ground-state
properties of finite nuclei.
We consider several different parametrizations of the RMF theory,
so that we can examine the model dependence of the results obtained.

This article is arranged as follows. In Sec.~\ref{sec:2},
we briefly describe the three methods used for the study of
neutron star crusts, namely, the TF approximation, the CP method,
and the CLD model with finite-size effects.
In Sec.~\ref{sec:3}, we discuss the RMF parameters
to be used in this study.
In Sec.~\ref{sec:4}, we show the numerical results and compare
the differences between these methods, as well as discuss
the effects of the symmetry energy on the neutron drip density
and properties of the inner crust.
Section~\ref{sec:5} is devoted to the conclusions.

\section{Model and methods}
\label{sec:2}

We employ the RMF theory to study a system consisting of
protons, neutrons, and electrons. In the RMF approach, nucleons interact via
the exchange of various mesons. The mesons considered are isoscalar scalar
and vector mesons ($\sigma$ and $\omega$) and the isovector vector meson ($\rho$).
Electrons and protons interact through the electromagnetic field $A^{\mu}$.
The Lagrangian density reads
\begin{eqnarray}
\label{eq:LRMF}
\mathcal{L}_{\rm{RMF}} & = & \sum_{i=p,n}\bar{\psi}_i
\left\{i\gamma_{\mu}\partial^{\mu}-\left(M+g_{\sigma}\sigma\right)
-\gamma_{\mu} \left[g_{\omega}\omega^{\mu} +\frac{g_{\rho}}{2}\tau_a\rho^{a\mu}
+\frac{e}{2}\left(1+\tau_3\right)A^{\mu}\right]\right\}\psi_i  \notag \\
& & +\bar{\psi}_{e}\left[i\gamma_{\mu}\partial^{\mu} -m_{e} +e \gamma_{\mu}
A^{\mu} \right]\psi_{e}  \notag \\
&& +\frac{1}{2}\partial_{\mu}\sigma\partial^{\mu}\sigma -\frac{1}{2}%
m^2_{\sigma}\sigma^2-\frac{1}{3}g_{2}\sigma^{3} -\frac{1}{4}g_{3}\sigma^{4}
\notag \\
&& -\frac{1}{4}W_{\mu\nu}W^{\mu\nu} +\frac{1}{2}m^2_{\omega}\omega_{\mu}%
\omega^{\mu} +\frac{1}{4}c_{3}\left(\omega_{\mu}\omega^{\mu}\right)^2  \notag
\\
&& -\frac{1}{4}R^a_{\mu\nu}R^{a\mu\nu} +\frac{1}{2}m^2_{\rho}\rho^a_{\mu}%
\rho^{a\mu} +\Lambda_{\rm{v}} \left(g_{\omega}^2
\omega_{\mu}\omega^{\mu}\right)
\left(g_{\rho}^2\rho^a_{\mu}\rho^{a\mu}\right) -\frac{1}{4}%
F_{\mu\nu}F^{\mu\nu},
\end{eqnarray}
where $W^{\mu\nu}$, $R^{a\mu\nu}$, and $F^{\mu\nu}$ are the antisymmetric
field tensors for $\omega^{\mu}$, $\rho^{a\mu}$, and $A^{\mu}$,
respectively. We include the $\omega$-$\rho$ coupling term as described
in~\cite{FSU}, which is essential in modifying the symmetry energy slope.
In the RMF approach, the meson fields are treated as
classical fields and the field operators are replaced by their expectation
values. For a static system, the nonvanishing expectation values are
$\sigma =\left\langle \sigma \right\rangle$, $\omega =\left\langle
\omega^{0}\right\rangle$, $\rho =\left\langle \rho^{30} \right\rangle$,
and $A =\left\langle A^{0}\right\rangle$. From the Lagrangian
density, we can derive the equations of motion for these mean fields
in a uniform or nonuniform system.

We employ the Wigner--Seitz approximation to describe the nonuniform matter
in neutron star crusts. In the present study, we focus on examining the
symmetry energy effects on properties of neutron star crusts around the
neutron drip density, where the inhomogeneous matter is composed of
spherical nuclei arranged in a body-centered-cubic (BCC) lattice.
Generally, nonspherical nuclei (pasta phases) may appear only
at densities higher than 0.05 fm$^{-3}$~\cite{Oyam07,Bao14}.
Therefore, we consider the matter of the crust to be divided
into spherical cells treated in the Wigner--Seitz approximation.
The Wigner--Seitz cell has the same volume as the unit cell in the BCC lattice.
The lattice constant $a$ and the Wigner--Seitz cell radius $r_{\rm{ws}}$ are
related to the cell
volume by $V_{\rm{cell}}=a^3=4 \pi r_{\rm{ws}}^3 / 3=N_b / n_b $, where $N_b$
and $n_{b}$ are the baryon number per cell and the average baryon number
density, respectively. We assume that each spherical nucleus is located in
the center of a charge-neutral cell consisting of a gas of nucleons and
electrons. It is well known that the electron screening effects are
negligible at subnuclear densities~\cite{Maru05}, so we ignore the
electron screening effect caused by the nonuniform charged particle
distributions and assume the electron density to be uniform inside the
Wigner--Seitz cell. At a given average baryon density $n_b$, the equilibrium
state is determined by minimizing the total energy density of the
system. To calculate the total energy per cell, we use the self-consistent
TF approximation with the RMF model, while the CP
method with Gibbs equilibrium conditions and the CLD model including
finite-size effects due to the surface and Coulomb
energies are adopted for comparison.

\subsection{Thomas--Fermi approximation}
\label{sec:2.1}

In the TF approximation, the total energy per cell can be written as
\begin{equation}
E_{\rm{cell}}=\int_{\rm{cell}}{\varepsilon }_{\rm{rmf}}(r)d^{3}r
  +{\varepsilon}_e V_{\rm{cell}} +\Delta E_{\rm{bcc}},
\label{eq:TFe}
\end{equation}%
where ${\varepsilon}_e$ denotes the electron kinetic energy density.
$\Delta E_{\rm{bcc}}$ is a correction term for the BCC
lattice, which is negligible when the nuclear size is much smaller
than the cell size~\cite{Oyam93,Shen11}.
${\varepsilon}_{\rm{rmf}}(r)$ is the local energy density at
radial position $r$, which is calculated in the RMF model as
\begin{eqnarray}
{\varepsilon }_{\rm{rmf}} &=&\displaystyle{\sum_{i=p,n}\frac{1}{\pi ^{2}}%
\int_{0}^{k_{F}^{i}}dk\,k^{2}\,\sqrt{k^{2}+{M^{\ast }}^{2}}}  \notag \\
&&+\frac{1}{2}(\nabla \sigma )^{2}+\frac{1}{2}m_{\sigma }^{2}\sigma ^{2}+%
\frac{1}{3}g_{2}\sigma ^{3}+\frac{1}{4}g_{3}\sigma ^{4}  \notag \\
&&-\frac{1}{2}(\nabla \omega )^{2}-\frac{1}{2}m_{\omega }^{2}\omega ^{2}-%
\frac{1}{4}c_{3}\omega ^{4}+g_{\omega }\omega \left( n_{p}+n_{n}\right)
\notag \\
&&-\frac{1}{2}(\nabla \rho )^{2}-\frac{1}{2}m_{\rho }^{2}\rho ^{2}
-\Lambda_{\rm{v}}g_{\omega }^{2}g_{\rho }^{2}\omega ^{2}\rho ^{2}+\frac{g_{\rho }}{2}%
\rho \left( n_{p}-n_{n}\right)   \notag  \\
&&-\frac{1}{2}(\nabla A)^{2}+eA\left( n_{p}-n_{e}\right) ,
\label{eq:ETF}
\end{eqnarray}%
where $n_{i}$ is the number density of species $i$ and
$M^{\ast}=M+g_{\sigma}\sigma $ is the effective nucleon mass.

From the Lagrangian density (\ref{eq:LRMF}), we obtain the
equations of motion for the mean fields:
\begin{eqnarray}
&&-\nabla ^{2}\sigma +m_{\sigma }^{2}\sigma +g_{2}\sigma ^{2}+g_{3}\sigma
^{3}=-g_{\sigma }\left( n_{p}^{s}+n_{n}^{s}\right) ,
\label{eq:eqms} \\
&&-\nabla ^{2}\omega +m_{\omega }^{2}\omega +c_{3}\omega^{3}
+2\Lambda_{\rm{v}}g^2_{\omega}g^2_{\rho}{\rho}^2 \omega
=g_{\omega}\left( n_{p}+n_{n}\right) ,
\label{eq:eqmw} \\
&&-\nabla ^{2}\rho +m_{\rho }^{2}{\rho}
+2\Lambda_{\rm{v}}g^2_{\omega}g^2_{\rho}{\omega}^2{\rho}
=\frac{g_{\rho }}{2}\left(n_{p}-n_{n}\right) ,
\label{eq:eqmr} \\
&&-\nabla ^{2}A=e\left( n_{p}-n_{e}\right) ,
\label{eq:eqma}
\end{eqnarray}%
where $n_{i}^{s}$ is the scalar density of species $i$. The equations of
motion for nucleons give the standard relations between the densities and
chemical potentials,
\begin{eqnarray}
\mu _{p} &=&{\sqrt{\left( k_{F}^{p}\right)^{2}+{M^{\ast }}^{2}}}+g_{\omega
}\omega +\frac{g_{\rho }}{2}\rho +e A,
\label{eq:mup} \\
\mu _{n} &=&{\sqrt{\left( k_{F}^{n}\right)^{2}+{M^{\ast }}^{2}}}+g_{\omega
}\omega -\frac{g_{\rho }}{2}\rho .
\label{eq:mun}
\end{eqnarray}%
We note that the chemical potential is spatially constant throughout the
Wigner--Seitz cell, while other quantities such as densities and mean fields
depend on the position $r$. In the Wigner--Seitz cell of neutron star crusts,
the conditions of $\beta$ equilibrium and charge neutrality are imposed,
which provide the constraints
\begin{eqnarray}
\mu _{n} &=&\mu _{p}+\mu _{e},
\label{eq:mue} \\
N_{e} &=&N_{p}=\int_{\rm{cell}}n_{p}(r)d^{3}r.
\label{eq:charge}
\end{eqnarray}

At a given average baryon density $n_{b}$, we minimize the total energy density with
respect to the cell radius $r_{\rm{ws}}$. To compute the total energy per cell at
fixed $r_{\rm{ws}}$ and $n_{b}$, we numerically solve the coupled
Eqs.~(\ref{eq:eqms})--(\ref{eq:eqma}) under the
constraints~(\ref{eq:mue}) and (\ref{eq:charge}).
In practice, we start with an initial guess for the mean fields
$\sigma (r)$, $\omega (r)$, $\rho (r)$, and $A(r)$, then determine the
chemical potentials $\mu _{n}$, $\mu _{p}$, and $\mu _{e}$
by the constraints~(\ref{eq:mue}) and (\ref{eq:charge})
and the given average density
$n_{b}=\left(N_{p}+N_{n}\right) /$ $V_{\rm{cell}}$.
Once the chemical potentials are obtained,
we can calculate various densities and solve
Eqs.~(\ref{eq:eqms})--(\ref{eq:eqma}) to get new mean fields.
This procedure is iterated until convergence is achieved.

\subsection{Coexisting phases method}
\label{sec:2.2}

In the CP method~\cite{Bao14,Mene08,Mene09,Maru05},
the matter inside the Wigner--Seitz cell
separates into a dense phase and a dilute phase with a sharp interface.
The coexisting phases satisfy Gibbs conditions for phase equilibrium,
which correspond to bulk equilibrium without finite-size effects.
The surface and Coulomb energies can be perturbatively taken into account
after the coexisting phases are achieved.
We denote the dense liquid phase and dilute gas phase
by $L$ and $G$, respectively. The Gibbs conditions for a nuclear
liquid phase in coexistence with a neutron gas at zero temperature
are written as
\begin{eqnarray}
\label{eq:CP1}
 & & P^{L} = P^{G},  \\
\label{eq:CP2}
 & & {\mu}^{L}_{n} = {\mu}^{G}_{n}.
\end{eqnarray}
The conditions of $\beta$ equilibrium and charge neutrality with a fixed
average baryon density $n_{b}$ provide the following constraints:
\begin{eqnarray}
\label{eq:CPbeta}
\mu_e &=& \mu_n^{L}-\mu_p^{L},  \\
\label{eq:CPcharge}
 n_e &=& n_p=u n_p^{L},\\
\label{eq:CPnb}
 n_b &=& u n_b^{L}+ \left(1-u\right) n_b^{G},
\end{eqnarray}
where $u$ denotes the volume fraction of the liquid phase.
We numerically solve Eqs.~(\ref{eq:CP1})--(\ref{eq:CPnb}) within the RMF model
to obtain all properties of the two coexisting phases and the volume fraction $u$
at given density $n_{b}$.

The total energy density of the system is given by
\begin{equation}
\label{eq:ews1}
\varepsilon=u{\varepsilon}_{\rm{bulk}}^{L}
            +\left(1-u\right){\varepsilon}_{\rm{bulk}}^{G}+{\varepsilon}_e
            +{\varepsilon}_{\rm{surf}}+{\varepsilon}_{\rm{Coul}},
\end{equation}
where ${\varepsilon}_{\rm{bulk}}^{L (G)}$ is the bulk energy density
of phase $L (G)$ obtained in the RMF model.
The surface and Coulomb energy densities for a spherical cell
are given by
\begin{eqnarray}
{\varepsilon}_{\rm{surf}} &=& \frac{3 \tau u}{r_d}, \label{eq:es1} \\
{\varepsilon}_{\rm{Coul}} &=& \frac{e^{2}}{5}
      \left(n^L_b Y^L_p\right)^{2}r_d^{2} u D\left( u\right) , \label{eq:ec1}
\end{eqnarray}%
with%
\begin{eqnarray}
\label{eq:eu1}
D\left( u\right) =1-\frac{3}{2}u^{1/3}+\frac{1}{2}u.
\end{eqnarray}%
Here $\tau$ is the surface tension, which can be obtained by
a TF calculation for semi-infinite nuclear matter~\cite{Mene10,Bao14,Cent98}.
$e=\sqrt{4\pi/137}$ is the electromagnetic coupling constant.
The radius of the droplet, $r_d$, is determined by minimizing
${\varepsilon}_{\rm{surf}}+{\varepsilon}_{\rm{Coul}}$, which leads to
${\varepsilon}_{\rm{surf}}=2{\varepsilon}_{\rm{Coul}}$.
The radius of the droplet and that of the Wigner--Seitz cell are, respectively,
given by
\begin{eqnarray}
\label{eq:RD}
r_d &=& \left[\frac{15\tau}{2 e^2
        \left(n^L_b Y^L_p\right)^2 D(u)}\right]^{1/3}, \\
\label{eq:RW}
r_{\rm{ws}} &=&   u^{-1/3} r_d.
\end{eqnarray}

We calculate the energy density of the cell by using Eq.~(\ref{eq:ews1})
at a given average baryon density $n_{b}$ and compare to that of corresponding
homogeneous phase. It is believed that the nonuniform matter in the Wigner--Seitz
approximation should have a smaller energy density than the homogeneous
phase at low density. However, we find that the energy density obtained
in the CP method is generally larger than that of the corresponding homogeneous
phase around the neutron drip density. The failure of the CP method
at low density may be due to the improper treatment of the surface
and Coulomb energies.

\subsection{Compressible liquid-drop model}
\label{sec:2.3}

In the CP method, the equilibrium conditions
are determined by the bulk properties without finite-size effects.
To incorporate the surface and Coulomb energies in determining the
equilibrium conditions, we employ the CLD model
to calculate the energy density of the Wigner--Seitz cell
and derive the equilibrium equations by minimization of the
total energy density including the surface and Coulomb
contributions~\cite{BBP71,Latt85,Latt91}.
The energy density of the cell is generally
expressed as a function of the following six variables:
the volume fraction and radius of the droplet ($u$ and $r_d$),
the baryon density and proton fraction inside the droplet ($n^L_{b}$ and $Y^L_{p}$),
and the number densities of the neutron and electron gases ($n^G_{b}$ and $n_e$).
The total energy density of the cell is given by
\begin{equation}
\label{eq:ews2}
\varepsilon=u {\varepsilon}_{\rm{bulk}}(n^L_{b},Y^L_{p})
            +\left({1-u}\right){\varepsilon}_{\rm{bulk}}(n^G_{b},0)
    +{\varepsilon}_e(n_e)
    +{\varepsilon}_{\rm{surf}}(u,r_d,\tau)
    +{\varepsilon}_{\rm{Coul}}(u,r_d,n^L_{b},Y^L_{p}),
\end{equation}
where ${\varepsilon}_{\rm{bulk}} (n^i_b,Y^i_p)$  is the energy density
of homogeneous nuclear matter in phase $i$ ($i=L,G$),
which can be calculated in the RMF model.
The surface and Coulomb terms are given by
Eqs.~(\ref{eq:es1}) and (\ref{eq:ec1}), respectively.
Under the constraints of charge neutrality
and fixed average baryon density given by Eqs.~(\ref{eq:CPcharge}) and~(\ref{eq:CPnb}),
there are only four independent variables and we may choose
$u$, $r_d$, $n^L_{b}$, and $Y^L_{p}$. Therefore, $n_e$ and $n^G_b$
are related to the independent variables by
\begin{eqnarray}
\label{eq:charge2}
 n_e &=& u n_b^{L} Y_b^L,\\
\label{eq:nb2}
 n_b^{G} &=& \frac{n_b-u n_b^{L}}{1-u}.
\end{eqnarray}

By minimizing the total energy density with respect to the independent
variables~\cite{Latt91}, we obtain the following equilibrium equations:
\begin{eqnarray}
\label{eq:ee21}
 0 =\frac{\partial \varepsilon}{\partial r_d} &:& \hspace{0.5cm}
  r_d =\left[\frac{15\tau}{2 e^2
        \left(n^L_b Y^L_p\right)^2 D(u)}\right]^{1/3}, \\
\label{eq:ee22}
0 =\frac{1}{un^L_b} \frac{\partial \varepsilon}{\partial Y_p^L} &:& \hspace{0.5cm}
  \mu_{e}=\mu_{n}^L-\mu_{p}^L-\frac{2 e^{2}}{5} n_b^L Y_p^L r_d^{2}D(u), \\
\label{eq:ee23}
0 =\frac{\partial \varepsilon}{\partial n^L_b}-\frac{Y_p^L}{n^L_b}
   \frac{\partial \varepsilon}{\partial Y_p^L} &:& \hspace{0.5cm}
  \mu_{n}^L = \mu_{n}^G, \\
\label{eq:ee24}
0 =\frac{n_b^L}{u}\frac{\partial \varepsilon}{\partial n^L_b}
                 -\frac{\partial \varepsilon}{\partial u} &:& \hspace{0.5cm}
P^L-P^G=\frac{ e^{2}}{5} \left(n^L_b Y^L_p\right)^2 r_d^{2}
        \left( 1-2u^{1/3}+u\right).
\end{eqnarray}
We note that the terms involving derivatives of the surface tension are ignored
in deriving these equilibrium equations.
As discussed by Iida and Oyamatsu~\cite{Oyam04},
the surface tension $\tau$ may depend
on the inner density and proton fraction ($n^L_{b}$ and $Y^L_{p}$).
Furthermore, $\tau$ could be affected by the size of the droplet ($r_d$),
which is known as a curvature correction to the surface tension~\cite{Haen00}.
However, the dependence of $\tau$ on these variables is poorly known,
especially in a neutron-rich system. This is because the surface tension
is generally obtained by a TF calculation for semi-infinite nuclear matter.
Due to the equilibrium conditions between the nuclear liquid and gas phases,
the surface tension would be a function of only one of the four variables
$n^L_{b}$, $Y^L_{p}$, $n^G_{b}$, and $Y^G_{p}$. Therefore, it is not
possible to obtain the partial derivatives of $\tau$ with respect to
each independent variable from this calculation.
For simplicity, we neglect contributions from the derivatives of
the surface tension in deriving the above equilibrium equations.
One can see that equilibrium equations~(\ref{eq:ee21})--(\ref{eq:ee24})
of the present paper are equivalent to Eqs.~(43)--(47) of Ref.~\cite{Cham08}.

By comparing Eq.~(\ref{eq:ee22}) with Eq.~(\ref{eq:CPbeta}), we can see that
the $\beta$ equilibrium condition is altered due to the inclusion of
finite-size effects in the minimization procedure.
The last term of Eq.~(\ref{eq:ee22}) comes from the Coulomb energy,
which favors a smaller electron chemical potential.
This leads to the conclusion that the electron fraction (equal to
the average proton fraction) is overestimated in the CP
method with bulk equilibrium.
Also, the inclusion of finite-size effects affects
the mechanical equilibrium as can be seen by comparing
Eq.~(\ref{eq:ee24}) with Eq.~(\ref{eq:CP1}).
The last term of Eq.~(\ref{eq:ee24}) comes from the sum of the surface
and Coulomb energies. Generally, the bulk pressure inside the droplet
is larger than that outside due to the surface and Coulomb contributions,
which leads to a higher density at the center of the droplet.

\section{Parameters}
\label{sec:3}

In this section, we discuss the choice of the RMF parameters
to be used in this study. The parameters of the RMF models are generally
fitted to nuclear matter saturation properties or ground-state
properties of finite nuclei. To study the properties of neutron star
crusts and compare the differences among various methods,
we consider four different RMF parametrizations,
NL3~\cite{NL3}, TM1~\cite{TM1}, FSU~\cite{FSU}, and IUFSU~\cite{IUFSU},
so that we can examine the model dependence of the results obtained.
These RMF models are known to be successful in reproducing
the ground state properties of finite nuclei including unstable ones.
The NL3 parametrization includes nonlinear terms of the $\sigma$ meson only,
while the TM1 parametrization includes nonlinear terms for both $\sigma$
and $\omega$ mesons. An additional $\omega$-$\rho$ coupling term is added
in the FSU and IUFSU parametrizations, and it plays an important role in
modifying the density dependence of the symmetry energy and
affecting the neutron star properties~\cite{Mene11,FSU,IUFSU,Horo01,Horo03,Prov13}.
The IUFSU parametrization was developed from FSU by reducing the neutron skin
thickness of $^{208}$Pb and increasing the maximum neutron star mass in
the parameter fitting~\cite{IUFSU}.
The TM1 model was successfully used to construct the equation of
state for supernova simulations and neutron star calculations~\cite{Shen02,Shen11}.
For completeness, we present the parameters and saturation properties
of these RMF models in Table~\ref{tab:1}.

\begin{figure}[htb]
\includegraphics[bb=65 288 511 676, width=8.6 cm,clip]{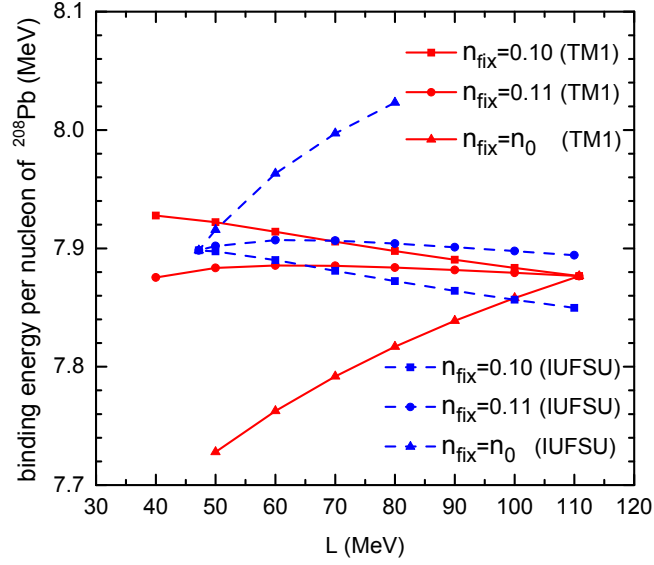}
\caption{(Color online) Binding energy per nucleon of $^{208}$Pb
vs the symmetry energy slope $L$ with different choices of $n_{\rm{fix}}$
based on the TM1 and IUFSU parametrizations.}
\label{fig:1Pb208}
\end{figure}

In order to examine the influence of the symmetry energy slope $L$,
we generate two sets of models based on the TM1 and IUFSU parametrizations.
We determine the model parameters by simultaneously adjusting
$g_{\rho}$ and ${\Lambda}_{\rm{v}}$ so as to achieve a given $L$ at saturation
density and keep $E_{\rm{sym}}$ fixed at a density of 0.11 fm$^{-3}$.
The choice of the fixed density $n_{\rm{fix}}=0.11\, \rm{fm}^{-3}$
is based on the following consideration.
In one set of generated models, the variation of $L$ at
saturation density would not affect the reproduction of
well-known properties of finite nuclei.
It has been pointed out that the binding energy of finite nuclei is essentially
determined by the symmetry energy at a density of $\sim 0.11\, \rm{fm}^{-3}$,
not by the symmetry energy at saturation density~\cite{Chen13,Horo01}.
To examine the sensitivity of the binding energy to the fixed density
$n_{\rm{fix}}$ of the symmetry energy, we perform a standard RMF calculation
as described in Refs.~\cite{Ring90,TM1}
for $^{208}$Pb using the two sets of generated models
with different choices of $n_{\rm{fix}}$.
One can see in Fig.~\ref{fig:1Pb208} that
the binding energy per nucleon of $^{208}$Pb remains almost unchanged
with varying $L$ using $n_{\rm{fix}}=0.11\, \rm{fm}^{-3}$,
whereas it deviates from the experimental value (7.87 MeV)
using $n_{\rm{fix}}=0.10\, \rm{fm}^{-3}$ or $n_{\rm{fix}}=n_0$
(where $n_0$ is the saturation density).
In Tables~\ref{tab:2} and~\ref{tab:3}, we present the parameters
$g_{\rho}$ and ${\Lambda}_{\rm{v}}$
generated based on TM1 and IUFSU by producing
a given $L$ at saturation density and fixed symmetry energy
at $n_{\rm{fix}}=0.11\, \rm{fm}^{-3}$.
We also show in these tables the symmetry energy at saturation density,
$E_{\rm{sym}} (n_0)$, and the neutron-skin thickness
$\Delta r_{np} =\langle r^2_n \rangle^{1/2} - \langle r^2_p \rangle^{1/2}$
of $^{208}$Pb, both of which generally increase with increasing $L$.
We stress that all models in each set have the same isoscalar saturation
properties and fixed symmetry energy at $n_{\rm{fix}}=0.11\, \rm{fm}^{-3}$,
but they have different symmetry energy slope $L$.
By using the set of models with different $L$, it is possible to study
the impact of $L$ on the neutron drip density and properties of
neutron star crusts.

\section{Results and discussion}
\label{sec:4}

In this section, we investigate the effects of the symmetry energy on
the neutron drip density and properties of neutron star crusts.
We first make a detailed comparison among the three methods used
for the study of neutron star crusts,
namely, the TF approximation, the CP method, and the CLD
model with finite-size effects. We analyze the differences among
these methods and explore their validity at low densities near
the neutron drip point. To study the influence of the symmetry
energy slope $L$, we employ the TF approximation, which is
considered to be self-consistent in the treatment of finite-size
effects and nucleon distributions.

\subsection{Comparison between different methods}
\label{sec:4.1}

To describe nonuniform matter in the Wigner--Seitz cell,
we consider three different methods: (1) the simple CP method
with bulk Gibbs equilibrium conditions; (2) the CLD model with
equilibrium conditions determined by including the surface and
Coulomb energies; and (3) the self-consistent TF approximation.
We note that treatments of surface and Coulomb energies
are obviously different among these methods. In the CP method,
Gibbs equilibrium conditions are used which correspond to bulk equilibrium
without finite-size effects, while the surface and Coulomb energies
are perturbatively incorporated after the two coexisting phases
are achieved. In the CLD model, equilibrium conditions are determined
by minimization of the total energy density including the surface
and Coulomb energies; therefore they are incorporated in a consistent
manner. In the TF approximation, the surface effect and nucleon
distributions are treated self-consistently, rather than
a sharp surface being assumed in the CP and CLD methods.
In addition, a neutron skin can be well described within
the TF approximation, but it is not explicitly included
in the CP and CLD methods.

\begin{figure}[htb]
\includegraphics[bb=49 316 499 716, width=8.6 cm,clip]{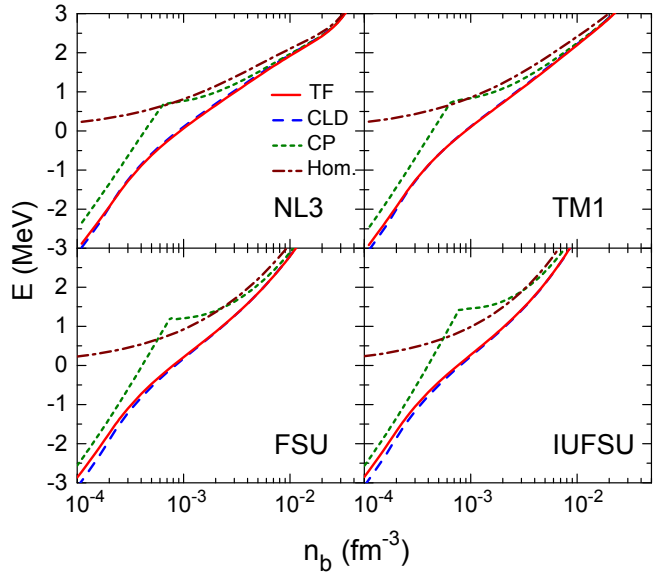}
\caption{(Color online) Energy per nucleon, $E$, as a function of
the average baryon density $n_b$ obtained using the TF, CLD,
and CP methods with different RMF parametrizations (NL3, TM1, FSU, and IUFSU).
For comparison, the results of homogeneous matter (Hom.) are also plotted.}
\label{fig:2ea}
\end{figure}

\begin{figure}[htb]
\includegraphics[bb=45 82 557 799, width=8.6 cm,clip]{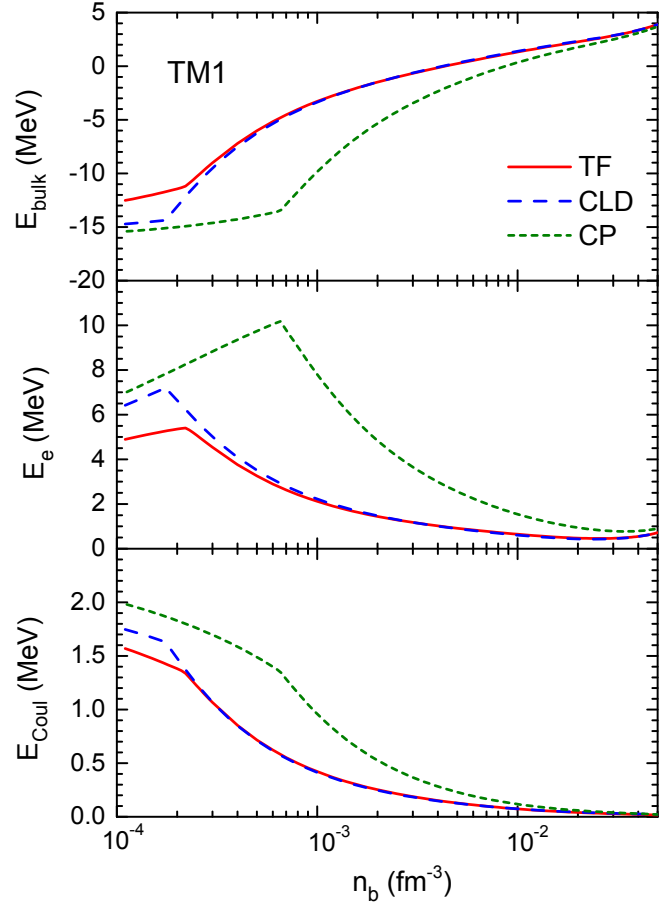}
\caption{(Color online) Comparison of the bulk energy per nucleon,
$E_{\rm{bulk}}$ (top panel), the electron kinetic energy per
nucleon, $E_{e}$ (middle panel), and the Coulomb energy per
nucleon, $E_{\rm{Coul}}$ (bottom panel) among the CP, CLD, and TF
methods using the TM1 parametrization.}
\label{fig:3ebec}
\end{figure}

\begin{figure}[htb]
\includegraphics[bb=34 316 499 716, width=8.6 cm,clip]{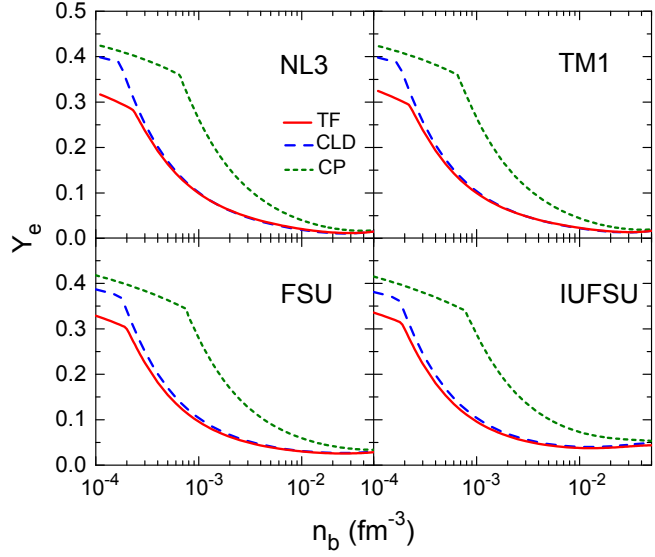}
\caption{(Color online) Electron fraction $Y_e$
as a function of $n_b$ obtained using the CP, CLD, and TF methods.}
\label{fig:4ye}
\end{figure}

In Fig.~\ref{fig:2ea}, we show the total energy per nucleon,
$E=\varepsilon /n_b -M$, as a function of the average baryon density $n_b$
obtained using the TF, CLD, and CP methods, while that of
homogeneous matter is also displayed.
It is interesting to see that the three methods yield very similar
$E$ at higher densities, but there are significant differences
at lower densities. Moreover, one can see that the simple CP method
fails to describe the nonuniform matter near the neutron drip
density, since $E$ of CP is larger than that of homogeneous matter.
We note that the kinks of CP at $n_b < 10^{-3}$ fm$^{-3}$ correspond
to the neutron drip point.
The failure of the CP method may be due to its improper treatment
of the surface and Coulomb energies.
It implies that the finite-size effect due to the surface and Coulomb
energies is too large to be treated perturbatively at low densities,
so that we have to include contributions from surface and Coulomb
energies in determining the equilibrium state as done in
the CLD and TF methods.
By comparing the results between CLD and CP,
one can see an obvious improvement due to the inclusion of
finite-size effects in the CLD method. Furthermore,
the results of CLD are very close to those obtained in
the self-consistent TF calculation.
In order to analyze the results of Fig.~\ref{fig:2ea}, we plot
various contributions to $E$ in Fig.~\ref{fig:3ebec}.
The Coulomb energy per nucleon,
$E_{\rm{Coul}}={\varepsilon}_{\rm{Coul}}/n_b$,
is calculated by using Eq.~(\ref{eq:ec1}) in the CP and CLD methods,
while it can be easily computed in the TF approximation by using
$E_{\rm{Coul}}=\frac{1}{2 N_b}
\int_{\rm{cell}} e A\left(r\right)\left[n_p\left(r\right)-n_e\right] d^3r$.
However, it is difficult to separate the surface energy from the bulk
energy in the TF approximation, because both are involved
in Eq.~(\ref{eq:ETF}). To estimate the surface energy in the TF approximation,
we use the equilibrium condition
${\varepsilon}_{\rm{surf}}=2\,{\varepsilon}_{\rm{Coul}}$
obtained in the liquid-drop model, which yields the sum
${\varepsilon}_{\rm{surf}}+{\varepsilon}_{\rm{Coul}}=3\,{\varepsilon}_{\rm{Coul}}$.
Therefore, we can define the bulk energy density in the TF approximation by
${\varepsilon}_{\rm{bulk}}=\left( E_{\rm{cell}}-\Delta E_{\rm{bcc}}\right)/V_{\rm{cell}}
-3{\varepsilon}_{\rm{Coul}}-{\varepsilon}_e$ according to Eq.~(\ref{eq:TFe}),
while it is given by ${\varepsilon}_{\rm{bulk}}=u{\varepsilon}_{\rm{bulk}}^{L}
+\left(1-u\right){\varepsilon}_{\rm{bulk}}^{G}$ in the CP and CLD methods.
In Fig.~\ref{fig:3ebec}, from top to bottom, we show, respectively,
the bulk energy per nucleon,
$E_{\rm{bulk}}={\varepsilon}_{\rm{bulk}}/n_b-M$,
the electron kinetic energy per nucleon,
$E_{e}={\varepsilon}_{e}/n_b$,
and the Coulomb energy per nucleon,
$E_{\rm{Coul}}$, obtained in the CP, CLD, and TF methods using the TM1 parametrization.
One can see that $E_e$ and $E_{\rm{Coul}}$ increase
with decreasing $n_b$, and the differences between CP and CLD methods
become very large at low density.
Due to the increasing contributions of $E_e$ and
$3E_{\rm{Coul}}$ (the sum of surface and Coulomb energies per nucleon),
the total energy per nucleon, $E$, obtained in the CP method is even larger
than that of homogeneous matter near the neutron drip density (see Fig.~\ref{fig:2ea}),
which implies that the simple CP method is not applicable to describing
nonuniform matter at low density.
In order to understand the differences in $E_e$ and $E_{\rm{Coul}}$ between
the CP and CLD methods, we display the electron fraction $Y_e=n_e/n_b$
as a function of $n_b$ in Fig.~\ref{fig:4ye}.
At a given $n_b$, a large $Y_e$ corresponds to large $n_e$ and $\mu_e$,
which results in more contributions from $E_e$ and $E_{\rm{Coul}}$.
One can see that $Y_e$ of the CP method is significantly larger than
that of the CLD and TF methods in all cases of Fig.~\ref{fig:4ye}.
This can be understood by comparing
Eqs.~(\ref{eq:CPbeta}) and~(\ref{eq:ee22}).
In the CP method, $\mu_e$ is determined by using Eq.~(\ref{eq:CPbeta}),
while an additional term (the last term) appears in Eq.~(\ref{eq:ee22})
caused by the Coulomb energy in the CLD method.
This term leads to a smaller $\mu_e$ in the CLD method compared to the CP case.
Therefore, we conclude that the inclusion of surface and Coulomb energies
in determining the equilibrium state plays a crucial role
in the description of nonuniform matter at low density.

\begin{figure}[htb]
\includegraphics[bb=32 312 499 716, width=8.6 cm,clip]{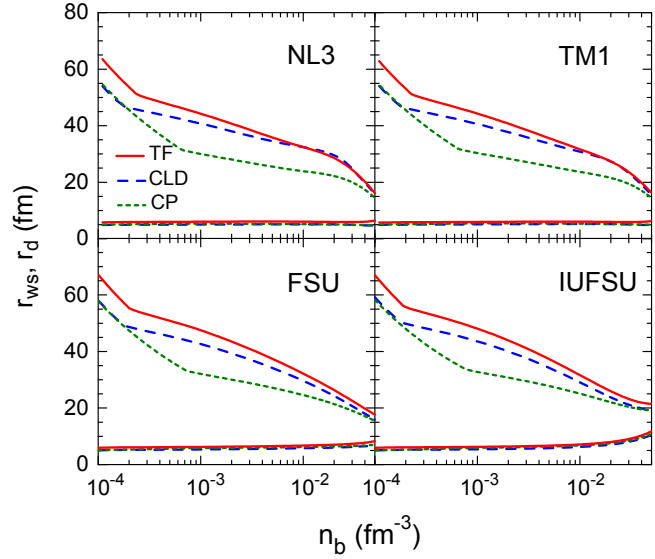}
\caption{(Color online) Wigner--Seitz cell radius $r_{\rm{ws}}$
and droplet radius $r_d$ as a function of $n_b$ obtained using
the TF, CLD, and CP methods. $r_d$ in the CP and CLD methods is
given in Eq.~(\ref{eq:RD}), while it is defined by
$r_d=\sqrt{\frac{5}{3}} \langle r^2_p \rangle^{1/2}$
in the TF approximation.}
\label{fig:5rcd}
\end{figure}

\begin{figure}[htb]
\includegraphics[bb=43 314 499 716, width=8.6 cm,clip]{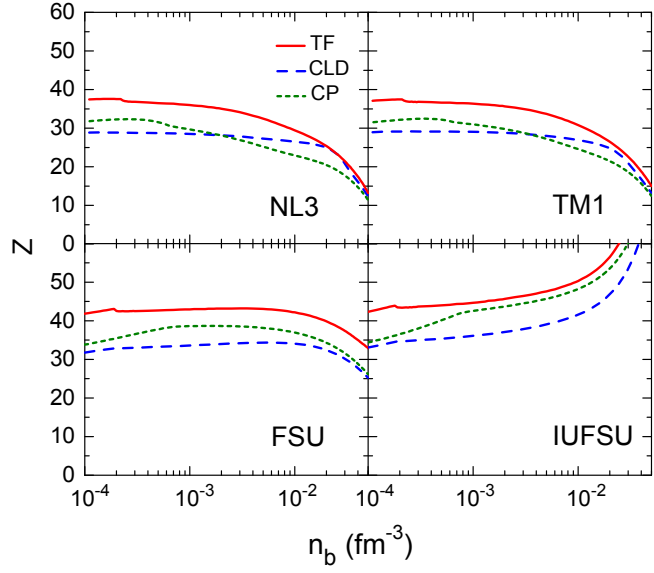}
\caption{(Color online) Proton number $Z$ of the droplet
as a function of $n_b$ obtained using the TF, CLD, and CP methods.}
\label{fig:6z}
\end{figure}

In Fig.~\ref{fig:5rcd}, we plot the radius of the droplet, $r_d$,
and that of the Wigner--Seitz cell, $r_{\rm{ws}}$,
as a function of $n_b$ obtained by using the TF, CLD, and CP methods.
In the CP and CLD methods, $r_d$ is given by Eq.~(\ref{eq:RD}),
while it is defined by
$r_d=\sqrt{\frac{5}{3}} \langle r^2_p \rangle^{1/2}$ in the TF approximation.
One can see that $r_d$ does not explicitly depend on $n_b$
and there is no significant difference among the three methods.
This is because the equilibrium nuclear size $r_d$ is mainly determined
by a competition between the surface and Coulomb energies,
which is a common feature in these methods.
On the other hand, $r_{\rm{ws}}$ obviously decreases with increasing $n_b$.
Moreover, $r_{\rm{ws}}$ in the CP method is generally smaller than that
of the CLD and TF methods.
This tendency is related to the behavior of $Y_e$ shown in Fig.~\ref{fig:4ye}.
As discussed above, a large $Y_e$ corresponds to large $n_e$ and $\mu_e$,
which results in a large volume fraction $u$ and a small $r_{\rm{ws}}$
according to the relations given in Eqs.~(\ref{eq:CPcharge}) and (\ref{eq:RW}).
In Fig.~\ref{fig:6z}, we present the proton number $Z$ of the droplet
as a function of $n_b$ obtained by using the TF, CLD, and CP methods.
It is well known that $Z$ is sensitive to the surface energy~\cite{Oyam07}.
We can see that the density dependence of $Z$ is relatively weak
at low density for all cases, while it shows a strong density dependence
with increasing $n_b$. The behavior of IUFSU is different from others
due to its relatively low value of $L$.
It has been shown in Refs.~\cite{Oyam07,Bao14} that
a small $L$ favors a large surface tension $\tau$,
which leads to a large $Z$ since $Z$ increases monotonically with $\tau$.
Comparing results among the three methods, we find that $Z$ of the TF method
is generally larger than that of the CP and CLD methods.
This may be due to the different treatment of nucleon distributions.
In the TF approximation, the surface effect and nucleon distributions
are calculated self-consistently and the neutron skin is well described.

\subsection{Neutron drip density}
\label{sec:4.2}

\begin{figure}[htb]
\includegraphics[bb=65 248 509 590, width=8.6 cm,clip]{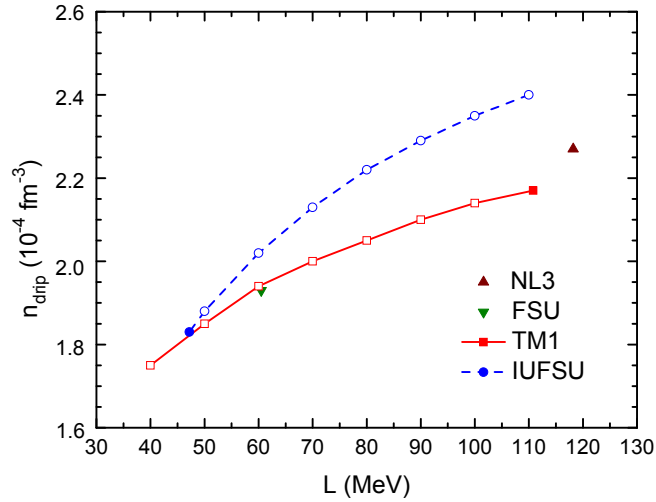}
\caption{(Color online) Neutron drip density $n_{\rm{drip}}$
as a function of $L$ using the two sets of models generated
from TM1 (red solid line with squares) and IUFSU (blue dashed line with circles).
The results obtained with the original TM1 and IUFSU models are
indicated by the filled square and filled circle, respectively.
The results of NL3 and FSU are represented by the up and down triangles.}
\label{fig:7ndrip}
\end{figure}

\begin{figure}[htb]
\includegraphics[bb=74 248 509 593, width=8.6 cm,clip]{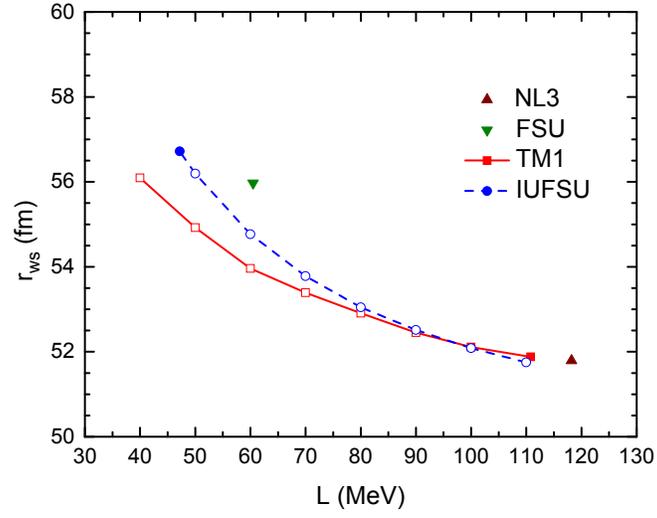}
\caption{(Color online) Same as Fig.~\ref{fig:7ndrip}, but for
the cell radius $r_{\rm{ws}}$ at the neutron drip density.}
\label{fig:8dripR}
\end{figure}

\begin{center}
\begin{figure}[thb]
\centering
\begin{tabular}{cc}
\includegraphics[bb=66 247 509 593, width=0.45\linewidth, clip]{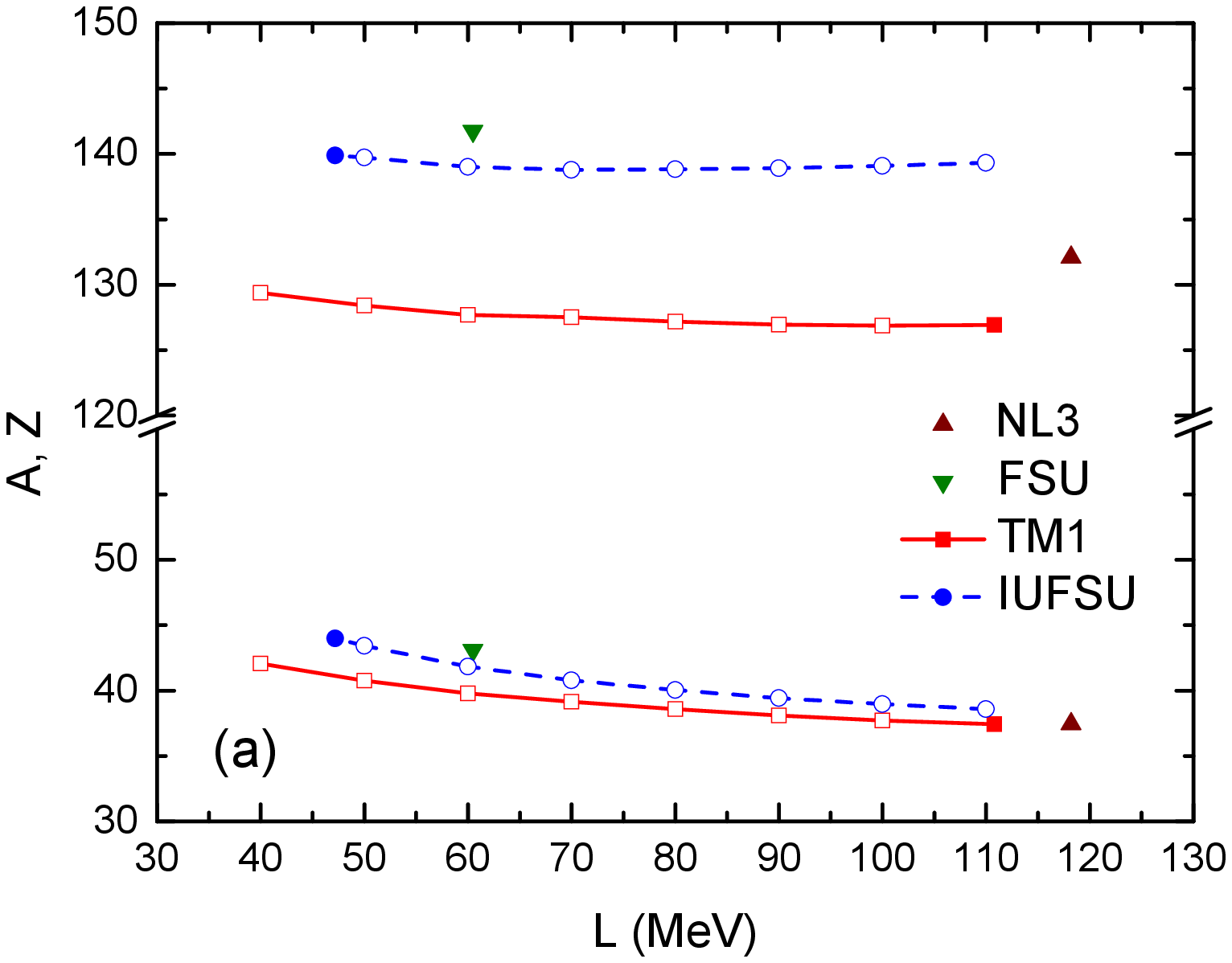}&
\includegraphics[bb=66 247 509 593, width=0.45\linewidth, clip]{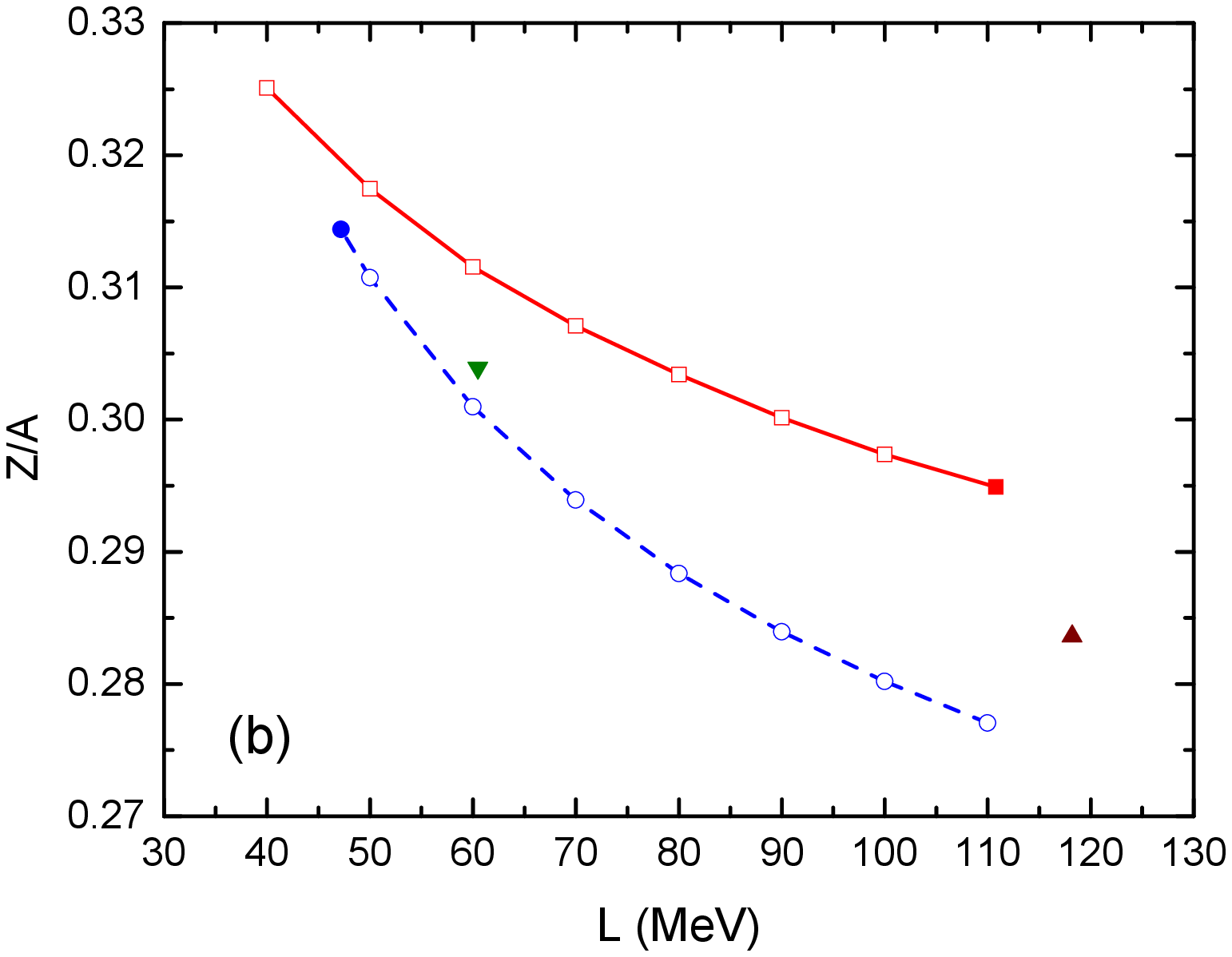}  \\
\includegraphics[bb=66 247 509 593, width=0.45\linewidth, clip]{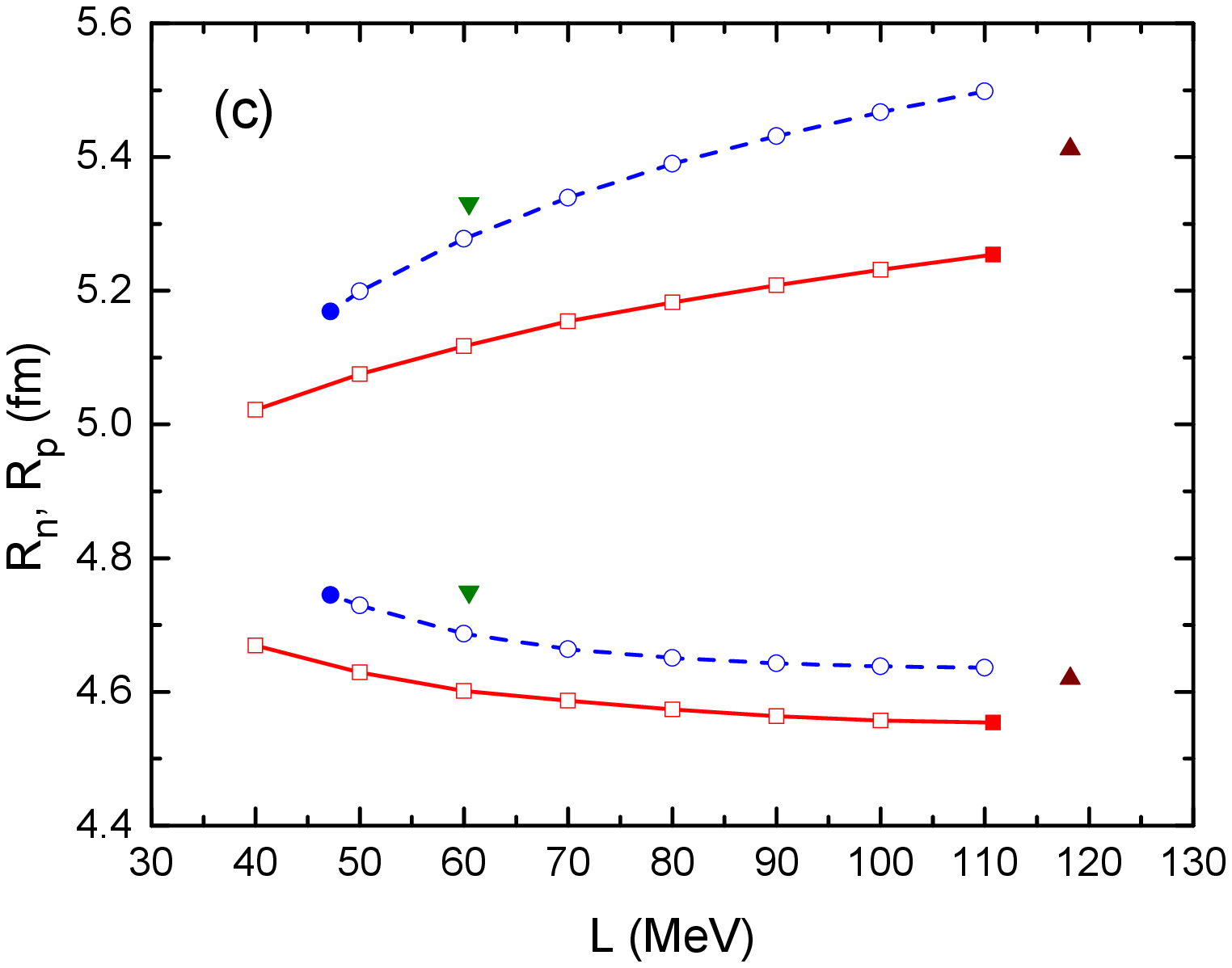}&
\includegraphics[bb=66 247 509 593, width=0.45\linewidth, clip]{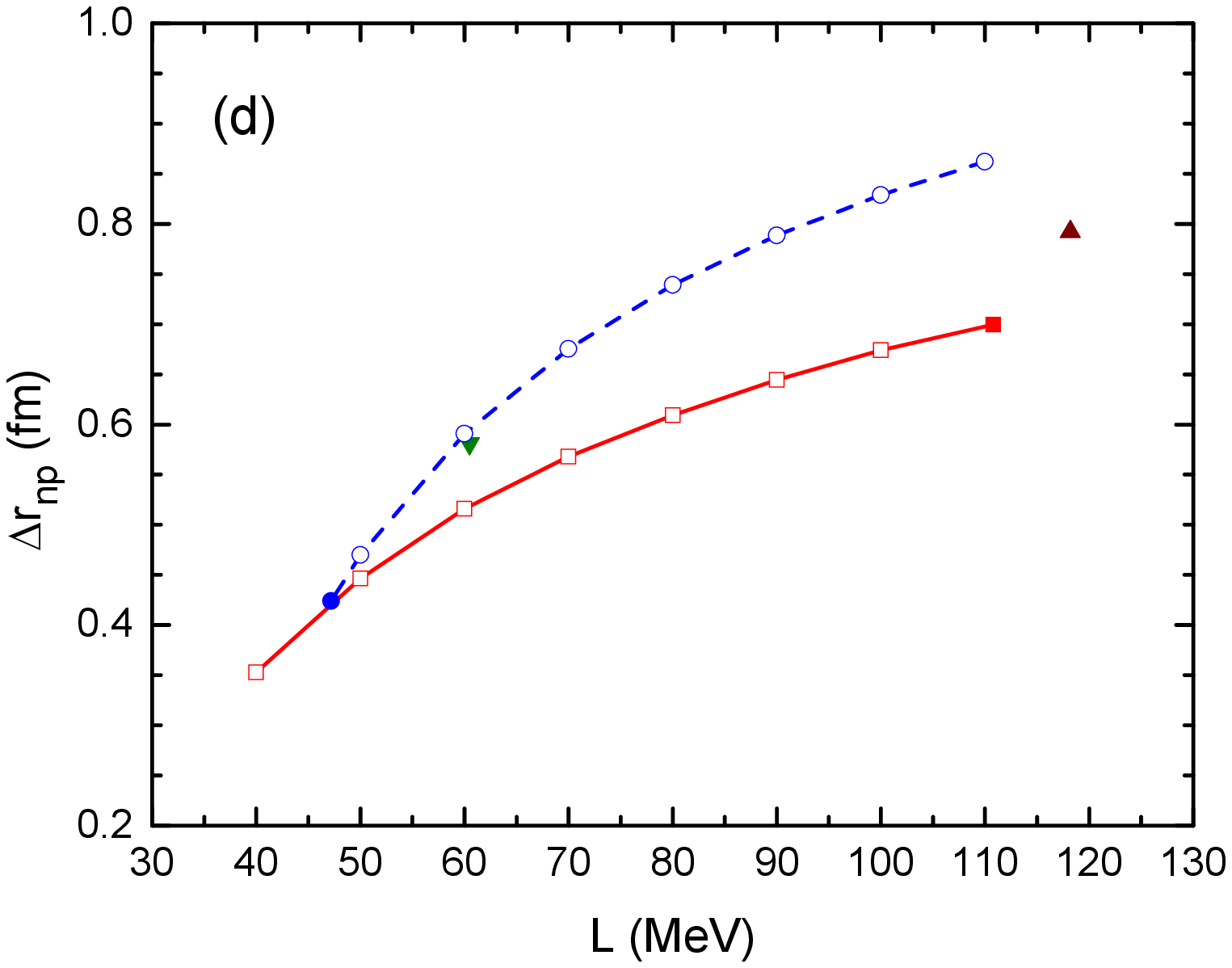} \\
\end{tabular}
\caption{(Color online) Properties of the nucleus at the neutron drip density
vs $L$ obtained in the TF calculation.
The nucleon number $A$ and the proton number $Z$ (a),
the proton fraction $Z/A$ (b), the rms radius of the neutron $R_n$
and that of the proton $R_p$ (c), and the neutron skin thickness
$\Delta r_{np}=R_n-R_p$ (d) are plotted.}
\label{fig:9dripA}
\end{figure}
\end{center}

\begin{figure}[htb]
\includegraphics[bb=48 248 505 592, width=8.6 cm,clip]{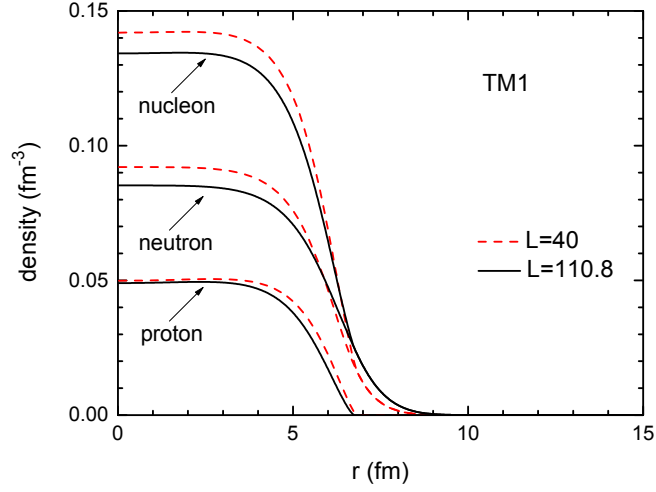}
\caption{(Color online) Nucleon density distributions in the Wigner--Seitz
cell at the neutron drip density obtained with $L=110.8$ MeV
(black solid lines) and $L=40$ MeV (red dashed lines) in the set of TM1.}
\label{fig:10dripD}
\end{figure}

We perform the self-consistent TF calculation to study the effects
of the symmetry energy on the neutron drip density. To examine the
influence of the symmetry energy slope $L$, we use two sets of models
generated from the TM1 and IUFSU parametrizations.
We note that all models in each set have the same isoscalar saturation
properties and fixed symmetry energy at $n_{\rm{fix}}=0.11\, \rm{fm}^{-3}$,
but they have different symmetry energy slope $L$.
The neutron drip point is determined by the condition $\mu_n = Mc^2$.
Beyond this point, neutrons begin to drip out of the nuclei and
form a free neutron gas.
In Fig.~\ref{fig:7ndrip}, we show the neutron drip density $n_{\rm{drip}}$
as a function of $L$ using the two sets of models generated from TM1 and IUFSU,
while the results of NL3 and FSU are also displayed.
It is found that $n_{\rm{drip}}$ increases with $L$
in both TM1 and IUFSU cases.
This tendency can be understood from the following analysis.
The neutron drip density is related to the nucleon number and radius of
the Wigner--Seitz cell as $n_{\rm{drip}}=A/ \frac{4}{3} \pi r_{\rm{ws}}^3 $.
The nucleon number $A$ at $n_{\rm{drip}}$ is not obviously affected
by $L$ [see Fig.~\ref{fig:9dripA}(a)].
However, the cell radius $r_{\rm{ws}}$ at $n_{\rm{drip}}$
decreases significantly with increasing $L$, as shown in Fig.~\ref{fig:8dripR}.
One reason for the decrease of $r_{\rm{ws}}$ is
because the generated models in each set have
fixed symmetry energy at $n_{\rm{fix}}=0.11\, \rm{fm}^{-3}$
with different $L$, and, therefore, a larger $L$ corresponds to
a larger symmetry energy $E_{\rm sym}$ near the saturation
density (see Tables~\ref{tab:2} and~\ref{tab:3}).
Based on the relation derived from the liquid-drop model,
$\mu_e = \mu_n^{L}-\mu_p^{L} \simeq 4 \delta E_{\rm{sym}}$
with $\delta=1-2Y^L_p$ being the neutron excess,
a large $E_{\rm{sym}}$ at the center of the nucleus
(corresponding to a large value of $L$) favors a high $\mu_e$,
although it corresponds to a small $\delta$ and a low nucleon density
in the center region (see Fig.~\ref{fig:10dripD}).
As mentioned above, a high value of $\mu_e$
results in a large volume fraction $u$ and a small $r_{\rm{ws}}$
according to the relations given in Eqs.~(\ref{eq:CPcharge}) and (\ref{eq:RW}).
Therefore, a larger $L$ in one set of generated models leads to
a smaller $r_{\rm{ws}}$ and a larger $n_{\rm{drip}}$, as shown
in Figs.~\ref{fig:7ndrip} and~\ref{fig:8dripR}.
The $L$ dependence of $n_{\rm{drip}}$ can also be explained by
the behavior of the neutron chemical potential $\mu_n$.
At the average baryon density $n_b$, a small $L$ generally
corresponds to a high $\mu_n$ due to the large contribution from
the $\rho$ meson [see Fig.~\ref{fig:13mu}(b)].
Therefore, the model with a smaller $L$ can reach the threshold condition for
the neutron drip $\mu_n = Mc^2$ at a lower density, which implies
an increasing $n_{\rm{drip}}$ with $L$, as shown in Fig.~\ref{fig:7ndrip}.

We display in Fig.~\ref{fig:9dripA} some properties of the nucleus
at the neutron drip density as a function of $L$
obtained in the TF calculation.
As one can see from Fig.~\ref{fig:9dripA}(a), the nucleon number
$A$ of the equilibrium nucleus is almost independent of $L$.
This is because the generated models with different $L$
have fixed symmetry energy at $n_{\rm{fix}}=0.11\, \rm{fm}^{-3}$,
which can produce very similar binding energies for finite nuclei
within one set of generated models (see Fig.~\ref{fig:1Pb208}).
The proton number $Z$ slightly decreases with increasing $L$,
which can be understood from
the $L$ dependence of the surface tension.
As discussed in Refs.~\cite{Oyam07,Bao14,Gril12},
a small $L$ favors a large surface tension $\tau$, which leads to
a large $Z$ since $Z$ increases monotonically with $\tau$.
The average proton fraction $Z/A$ of the nucleus is found to decrease
with increasing $L$ [see Fig.~\ref{fig:9dripA}(b)],
which is caused by the decrease of $Z$ with $L$.
The root-mean-square (rms) radius of the neutron ($R_n$) increases
with $L$, whereas that of the proton ($R_p$)
decreases [see Fig.~\ref{fig:9dripA}(c)].
The difference between $R_n$ and $R_p$,
known as the neutron skin thickness ($\Delta r_{np}$),
is displayed in Fig.~\ref{fig:9dripA}(d).
It is well known that a larger $L$ results in a thicker neutron
skin~\cite{IUFSU,Horo01,Horo03,Gril12,Cent10},
which is also observed in Fig.~\ref{fig:9dripA}(d).

We plot in Fig.~\ref{fig:10dripD} the nucleon density distributions
in the Wigner--Seitz cell at the neutron drip density obtained
with two extreme values of $L$ in the set of TM1.
It is shown that the nucleon distributions, especially the neutron
distributions, can be significantly affected by the value of $L$.
One can see that a large $L$ results in a small nucleon density
at the center of the cell. This may be understood from the
analysis based on a liquid-drop model. As discussed
by Iida and Oyamatsu~\cite{Oyam04},  the equilibrium density
of the nucleon liquid can be estimated by the condition
of zero pressure when there is no neutron gas outside.
The bulk pressure vanishes at
$n^L_b=n_0-\frac{3L n_0}{K} \delta^2$
derived from the liquid-drop model~\cite{Oyam04}.
This implies that the equilibrium density $n^L_b$ decreases
with increasing $L$ for a fixed neutron excess $\delta$.
On the other hand, a larger $L$ in one set of generated models
corresponds to a larger symmetry energy $E_{\rm sym}$ near the
saturation density, as mentioned above.
As a result, a larger $L$ favors fewer neutrons (equivalent
to a smaller $\delta$) in the central region of the nucleus due to
its larger $E_{\rm sym}$. Meanwhile, more neutrons are distributed
in the surface region due to its smaller $E_{\rm sym}$ for a larger $L$
at very low density. Therefore, a large value of $L$ results
in relatively large neutron rms radius and neutron skin thickness
[see Figs.~\ref{fig:9dripA}(c) and~\ref{fig:9dripA}(d)].

\subsection{Properties of nuclei in neutron star crusts}
\label{sec:4.3}

\begin{center}
\begin{figure}[thb]
\centering
\begin{tabular}{ccc}
\includegraphics[bb=24 159 471 798, width=0.32\linewidth, clip]{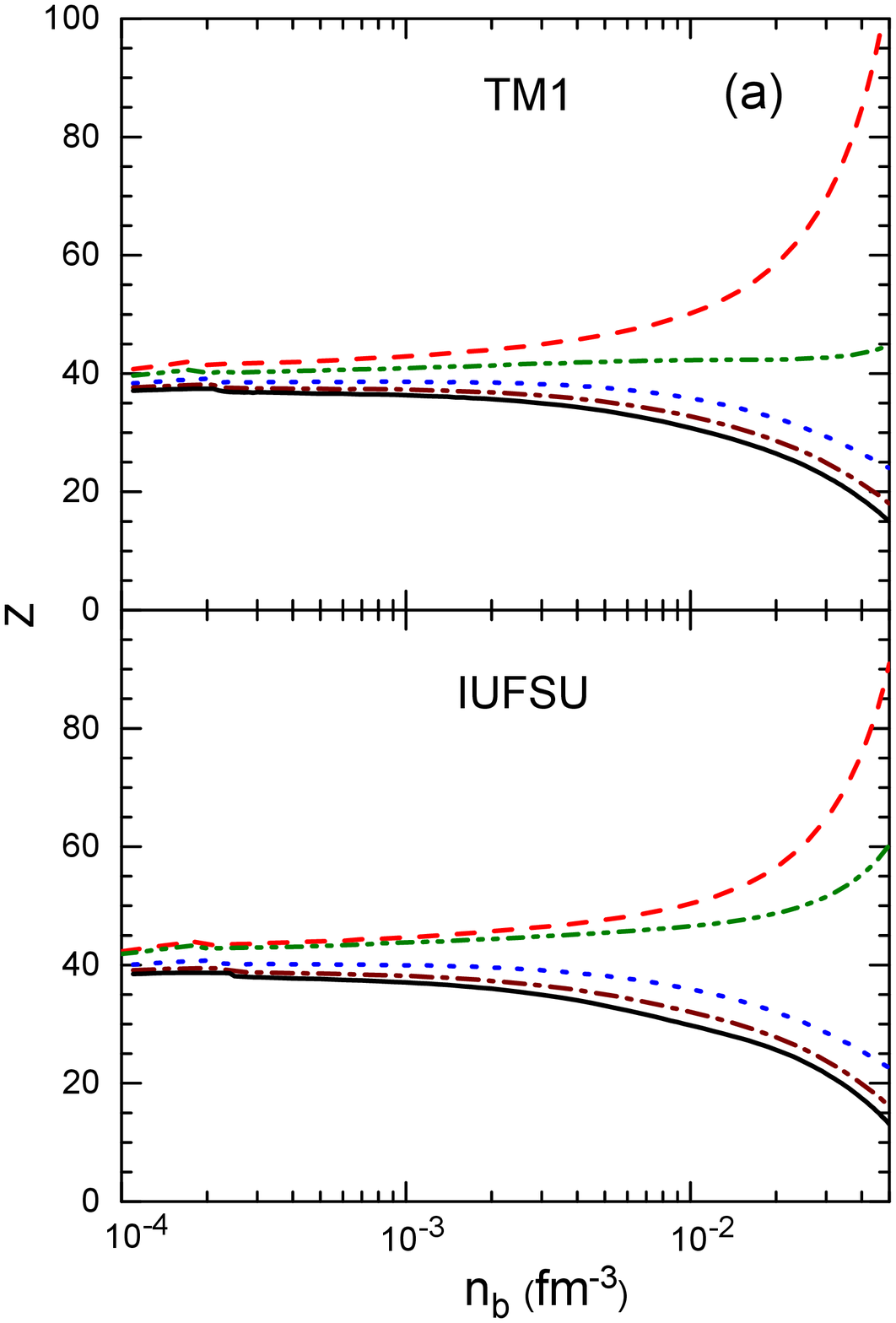}&
\includegraphics[bb=24 159 471 798, width=0.32\linewidth, clip]{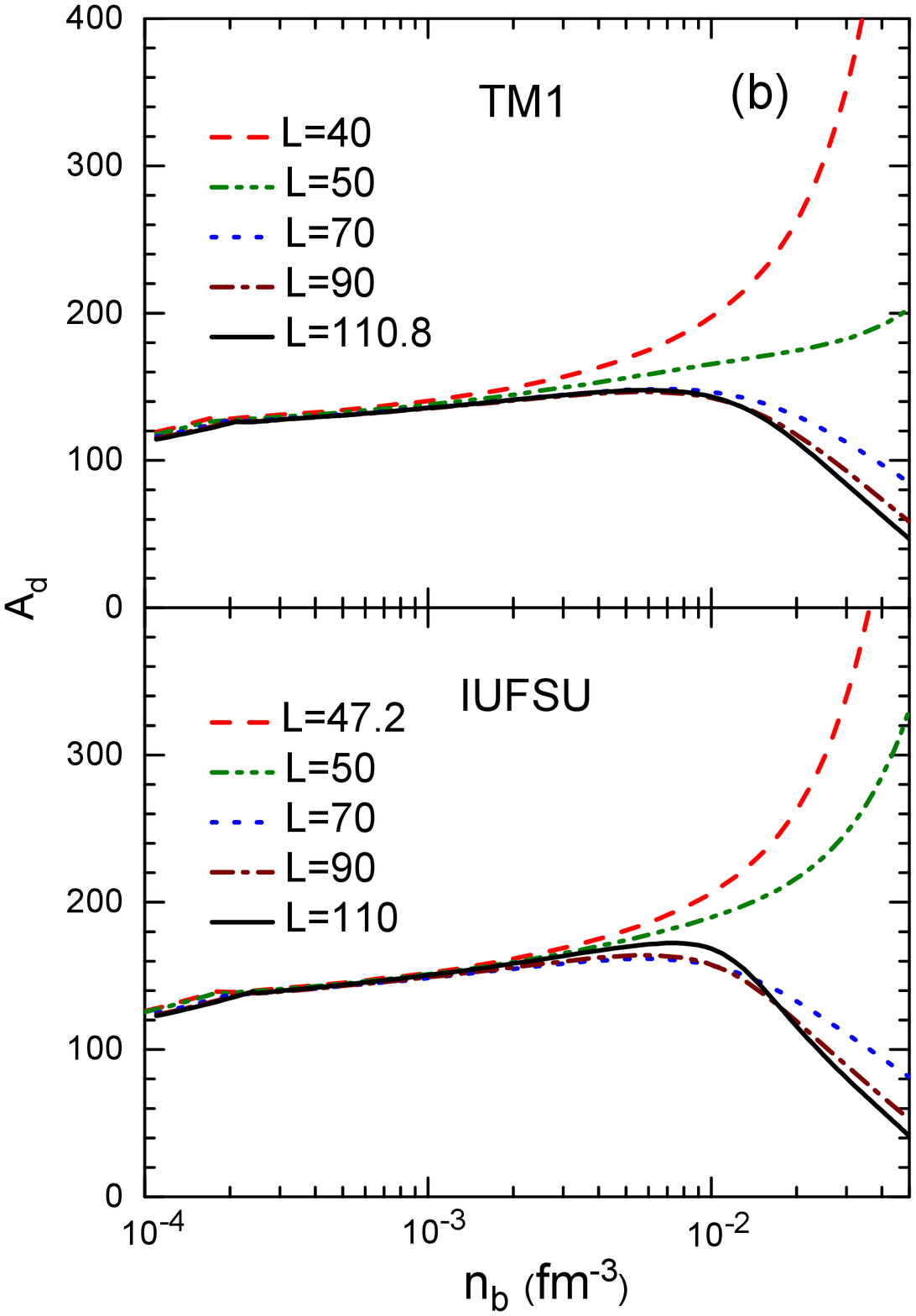}&
\includegraphics[bb=24 159 471 798, width=0.32\linewidth, clip]{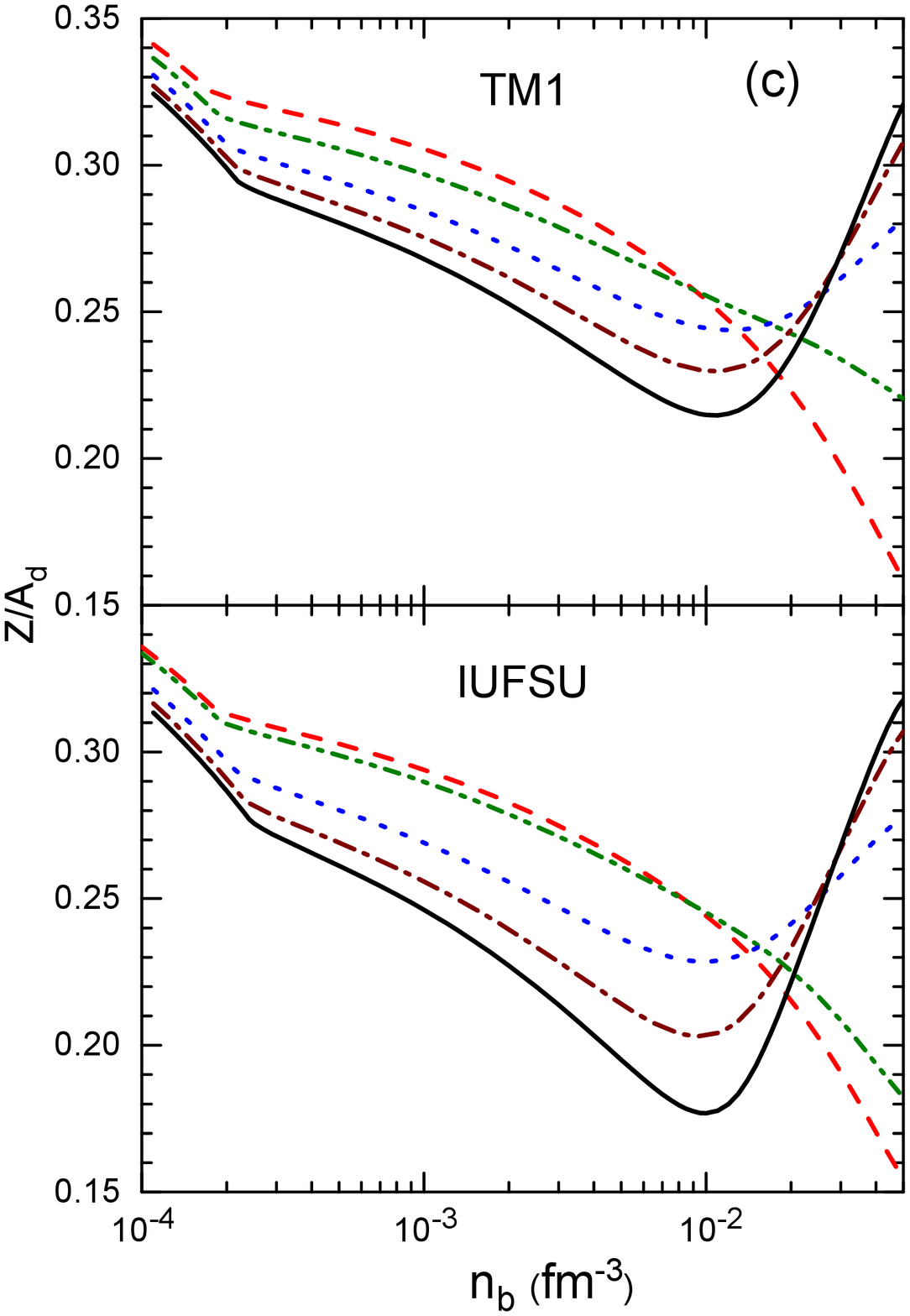}  \\
\end{tabular}
\caption{(Color online) Properties of the nucleus in neutron star crusts, such as
the proton number $Z$ (a), the nucleon number $A_d$ (b),
and the proton fraction $Z/A_d$ (c), as a function
of $n_b$ obtained in the TF approximation using the two sets
of generated models.}
\label{fig:11tfdrop}
\end{figure}
\end{center}

We employ the TF approximation to study the effects of the symmetry
energy on properties of nuclei in the inner crust.
Above the neutron drip density $n_{\rm{drip}}$,
a gas of free neutrons coexists with
a lattice of spherical nuclei, and the equilibrium nuclei
become more and more neutron rich as the density increases.
In Fig.~\ref{fig:11tfdrop}, we display the droplet proton number $Z$,
nucleon number $A_d$, and proton fraction $Z/A_d$
as a function of the average baryon density $n_b$ using the two sets of
generated models. The droplet nucleon number $A_d$ is defined by subtracting
the background neutrons in order to isolate the nucleus from a
surrounding neutron gas~\cite{Gril12,Sub02}.
It is shown that $Z$ and $A_d$ weakly depend on $n_b$ at lower densities,
while they rapidly change at relatively high densities.
For the $L$ dependence of $Z$ and $A_d$, it is found that $Z$
decreases monotonically with increasing $L$, while $A_d$ is almost
independent of $L$ at low densities.
These behaviors are consistent with those shown in Fig.~\ref{fig:9dripA}(a).
Our results are very similar to those reported in Ref.~\cite{Gril12}.
The $L$ dependence of $Z$ may be understood from the behavior of the surface
tension $\tau$. Based on the size equilibrium condition of the liquid-drop model,
${\varepsilon}_{\rm{surf}}=2\,{\varepsilon}_{\rm{Coul}}$,
a large value of $\tau$ leads to large nuclear size $r_d$ and proton number $Z$.
It has been shown in Refs.~\cite{Oyam07,Bao14,Gril12} that
a large $L$ corresponds to a small $\tau$. Therefore,
a small $Z$ is achieved for a large $L$ due to its small $\tau$.
The $L$ dependence of $A_d$ at high densities is mainly because
the nuclear size increases with decreasing $L$ (equivalent to increasing $\tau$).
The proton fraction $Z/A_d$ at low densities is found to decrease
with increasing $L$, which is related to the behaviors of $Z$ and $A_d$,
but the opposite tendency is observed at high densities.
A similar behavior of $Z/A_d$ was also observed in Fig.~4(f) of Ref.~\cite{Gril12}.
The strong $L$ dependence at high densities obtained in the present TF
calculation is consistent with that shown in our previous study
using the CP method~\cite{Bao14}, where a large value of $L$
leads to small $\tau$, $Z$, $A_d$, and $r_d$ values
(see Figs.~4--7 of Ref.~\cite{Bao14}).
It has been shown in Sec.~\ref{sec:4.1} that the difference
between the TF and CP methods is relatively small in the high-density region.

\begin{center}
\begin{figure}[thb]
\centering
\begin{tabular}{ccc}
\includegraphics[bb=13 154 471 798, width=0.32\linewidth, clip]{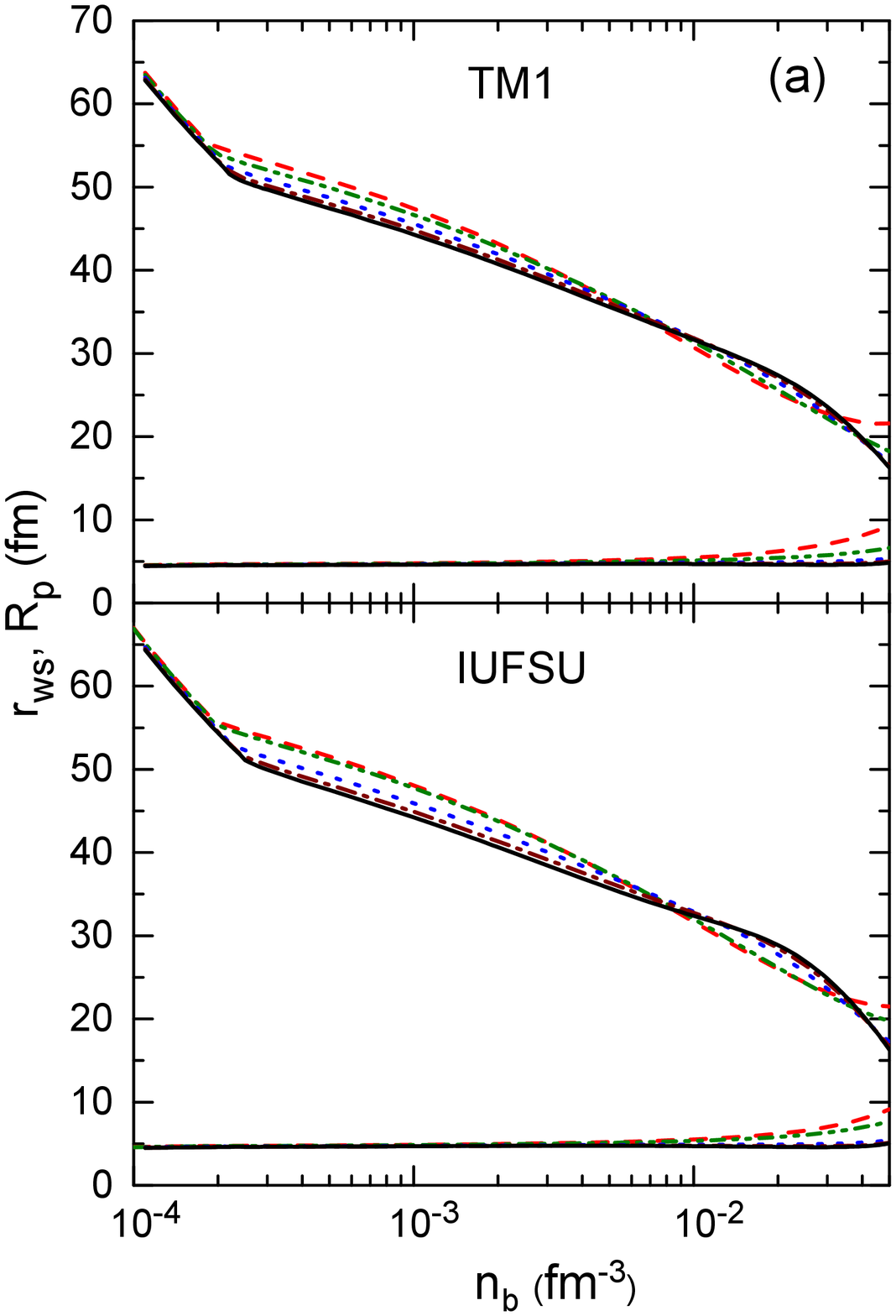}&
\includegraphics[bb=13 154 471 798, width=0.32\linewidth, clip]{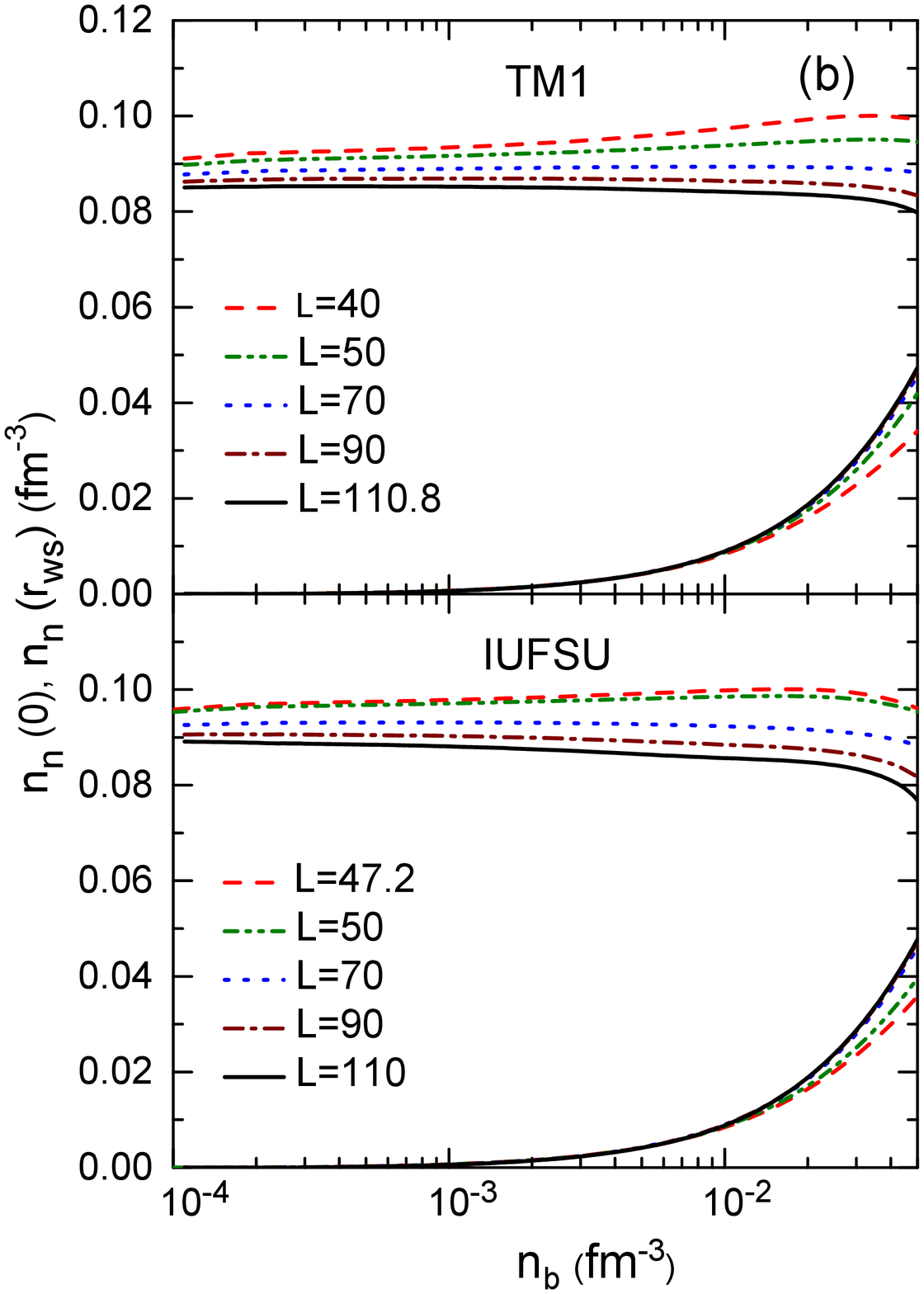}&
\includegraphics[bb=13 154 471 798, width=0.32\linewidth, clip]{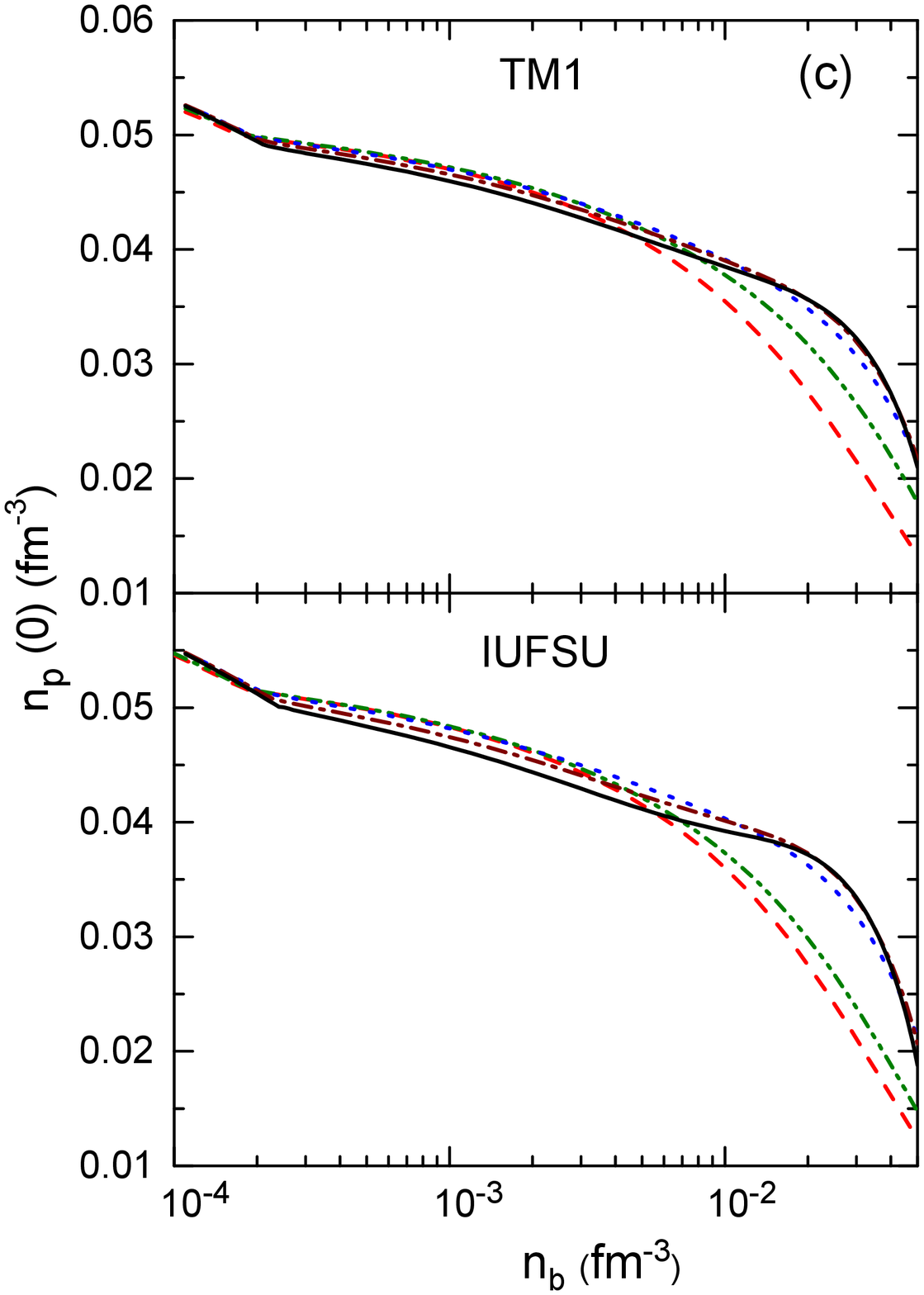} \\
\end{tabular}
\caption{(Color online) Equilibrium properties of the Wigner--Seitz cell
as a function of $n_b$ obtained in the TF approximation
using the two sets of generated models.
The cell radius $r_{\rm{ws}}$ and the proton rms radius $R_p$ (a),
the neutron density at the center, $n_{n}(0)$, and that at the boundary,
$n_{n}(r_{\rm{ws}})$ (b), and the proton density at the center, $n_{p}(0)$ (c),
are plotted.}
\label{fig:12ws}
\end{figure}
\end{center}

\begin{center}
\begin{figure}[thb]
\centering
\begin{tabular}{ccc}
\includegraphics[bb=22 150 471 798, width=0.32\linewidth, clip]{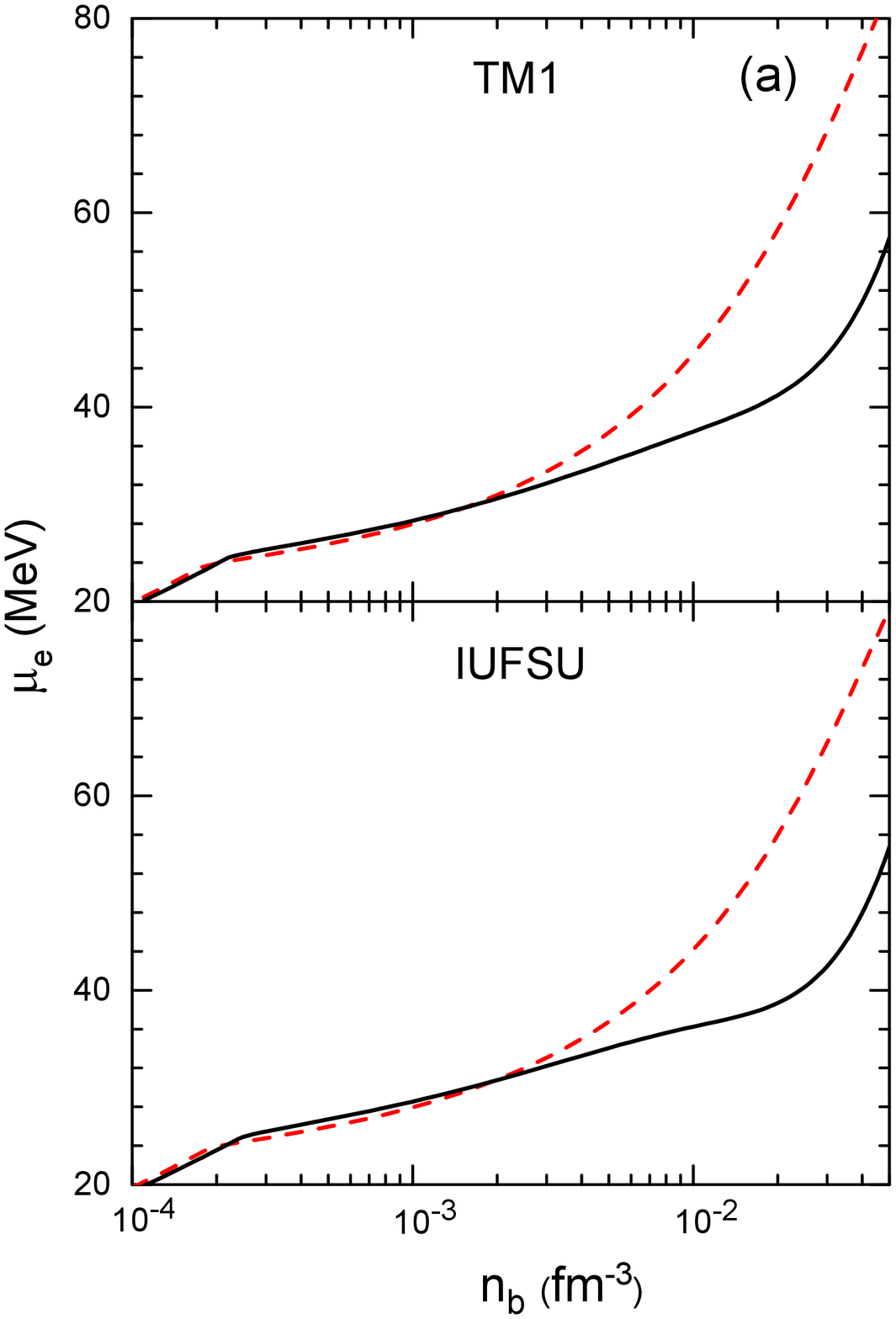}&
\includegraphics[bb=22 150 471 798, width=0.32\linewidth, clip]{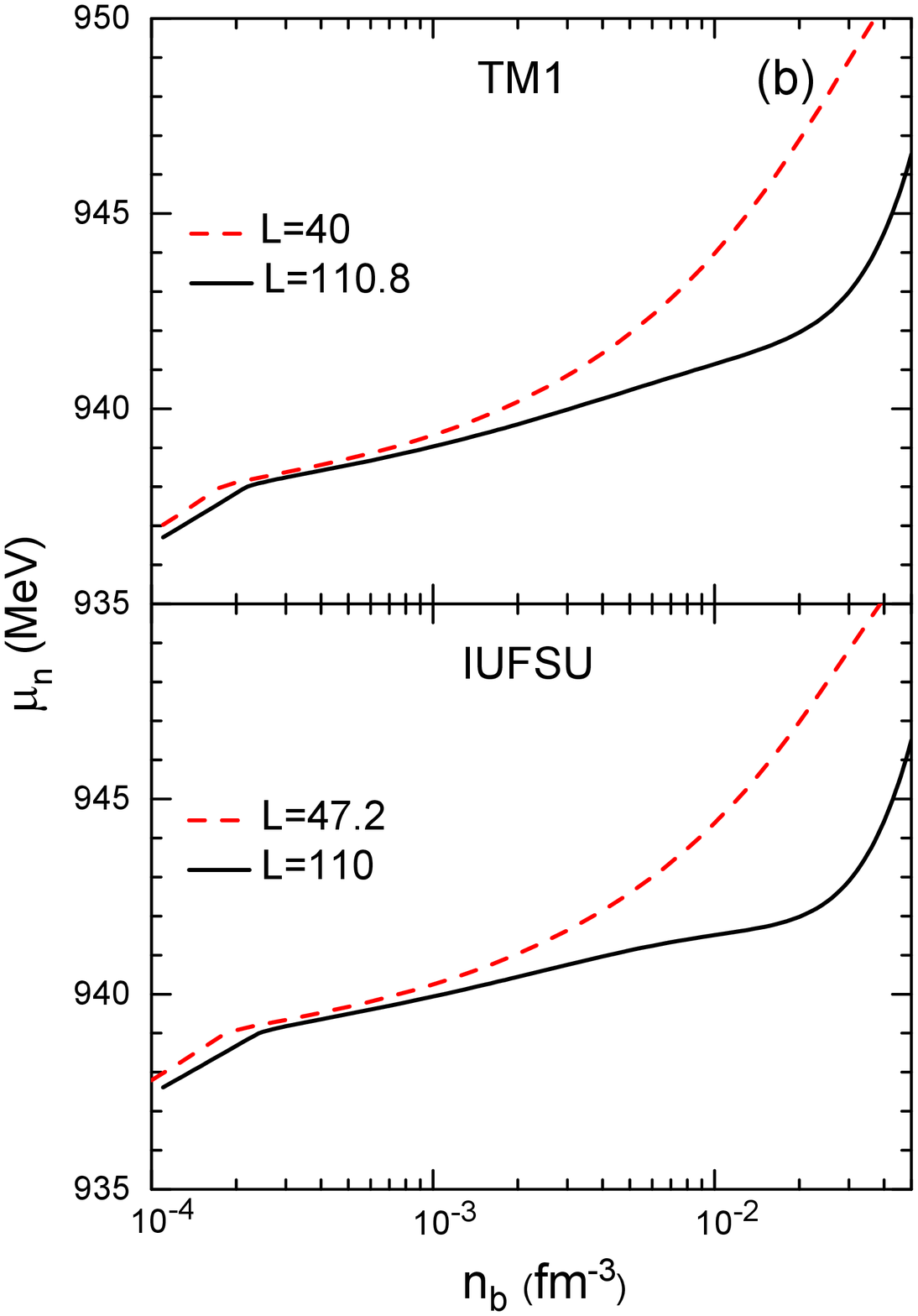}&
\includegraphics[bb=22 150 471 798, width=0.32\linewidth, clip]{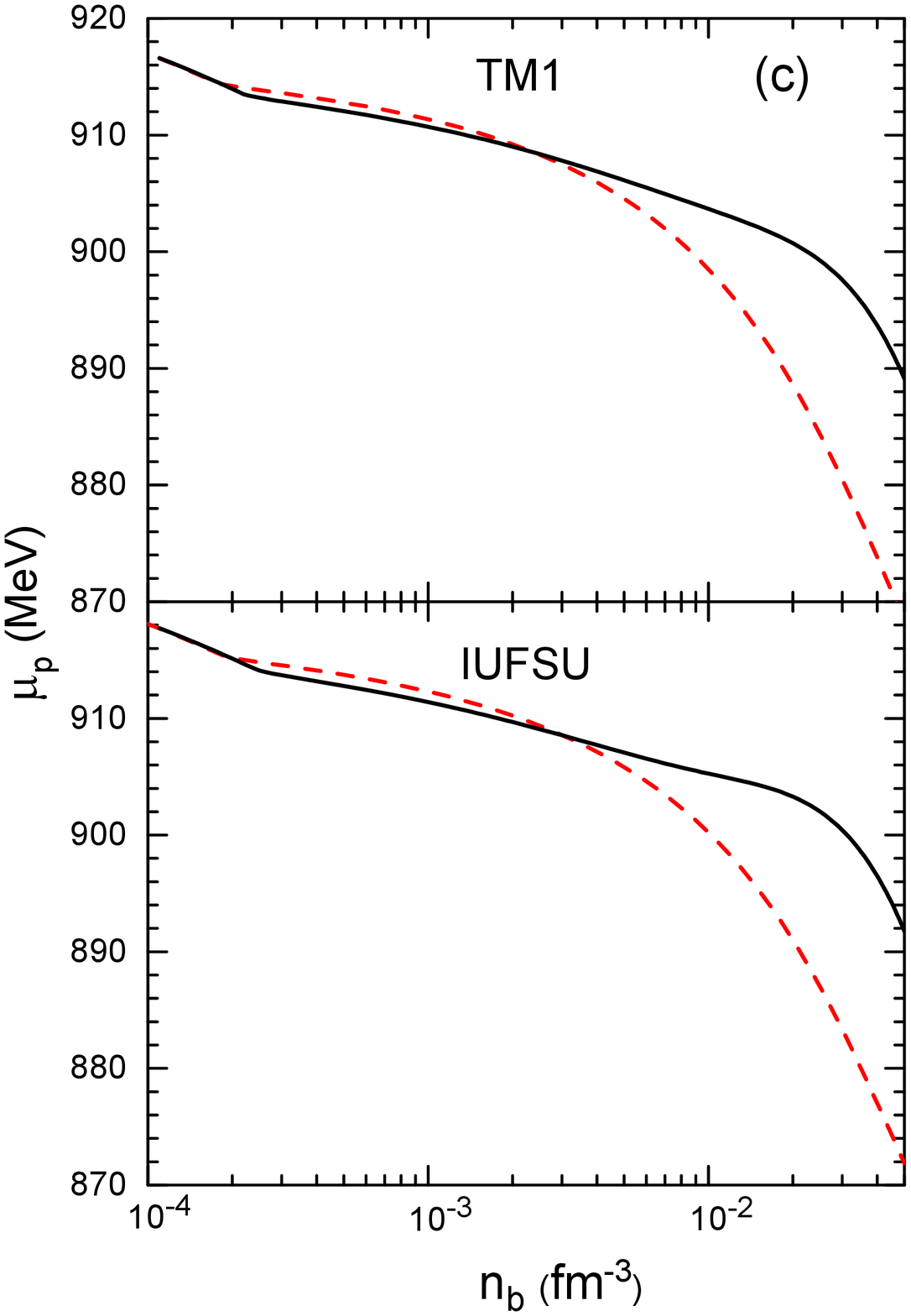} \\
\end{tabular}
\caption{(Color online) Chemical potentials of electrons, $\mu_e$ (a),
neutrons, $\mu_n$ (b), and protons, $\mu_p$ (c), as a function
of $n_b$ obtained in the TF approximation with the smallest
and largest values of $L$ in the two sets of generated models.}
\label{fig:13mu}
\end{figure}
\end{center}

\begin{figure}[htb]
\includegraphics[bb=30 62 566 761, width=8.6 cm,clip]{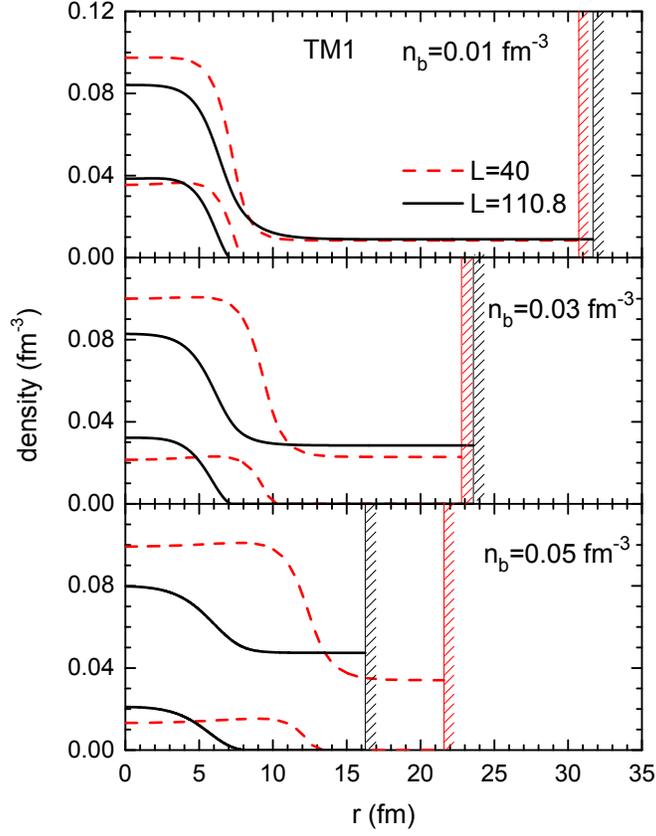}
\caption{(Color online) Density distributions of neutrons (upper curves)
and protons (lower curves) in the Wigner--Seitz cell
at average baryon densities $n_b=0.01,\,0.03,$ and $0.05\,\rm{fm}^{-3}$
(top to bottom) obtained with $L=110.8$ MeV (black solid lines)
and $L=40$ MeV (red dashed lines) in the set of TM1.
The cell radius $r_{\rm{ws}}$ is indicated by the hatching.}
\label{fig:14wsD}
\end{figure}

In Figs.~\ref{fig:12ws} and~\ref{fig:13mu}, we present equilibrium
properties of the Wigner--Seitz cell as a function of $n_b$ obtained
in the TF approximation using the two sets of generated models.
For clarity of presentation, we show chemical potentials $\mu_e$,
$\mu_n$, and $\mu_p$ in Fig.~\ref{fig:13mu} with only
the smallest and largest values of $L$ in each set of generated models.
One can see from Fig.~\ref{fig:12ws}(a) that the radius of
the Wigner--Seitz cell, $r_{\rm{ws}}$,
significantly decreases with increasing $n_b$, while the proton
rms radius $R_p$ weakly depends on $n_b$ only at high densities.
These behaviors are consistent with those shown
in Fig.~\ref{fig:5rcd}, where the droplet radius $r_d$ in the TF
approximation is calculated from the proton rms radius $R_p$ as
$r_d=\sqrt{\frac{5}{3}} R_p$.
The decrease of $r_{\rm{ws}}$ is caused by the
increase of nuclear volume fraction $u$ with increasing $n_b$.
On the other hand, the proton density at the center of the cell, $n_p(0)$,
obviously decreases with increasing $n_b$ [see Fig.~\ref{fig:12ws}(c)].
This is because the matter gets more neutron rich and the difference
between the neutron and proton chemical potentials,
which is equivalent to the electron chemical potential as
$\mu_e = \mu_n-\mu_p$, becomes larger as the density
increases [see Fig.~\ref{fig:13mu}(a)].
Moreover, the decrease of $n_p(0)$ at high
densities shows a strong $L$ dependence; namely, a small $L$ leads to
a rapid decrease of $n_p(0)$. This may be understood from
the influence of the $\omega$-$\rho$ coupling term,
which plays an important role in neutron-rich matter.
At the center of the cell, we have the following relation between
densities and chemical potentials according to
Eqs.~(\ref{eq:eqmr})--(\ref{eq:mue}):
\begin{eqnarray}
\mu_e = \mu_n-\mu_p &=& \sqrt{\left( k_{F}^{n}\right)^{2}+{M^{\ast}}^{2}}
                       -\sqrt{\left( k_{F}^{p}\right)^{2}+{M^{\ast}}^{2}}
                     -eA -g_{\rho}\rho  \nonumber \\
  & \simeq & \frac{(3\pi^2)^{2/3}}{2M^{\ast}}\left( n_{n}^{2/3}-n_{p}^{2/3}\right)
  -eA +\frac{g_{\rho }^{2}}{2\left(m_{\rho }^{2}
+2\Lambda_{\rm{v}}g_{\omega}^{2}g_{\rho}^{2}{\omega}^{2}\right) }\left( n_{n}-n_{p}\right).
\label{eq:cwr}
\end{eqnarray}
As $n_b$ increases, $\mu_e=\mu_n-\mu_p$ increases monotonically,
as shown in Fig.~\ref{fig:13mu}(a), which yields increasing $n_{n}(0)-n_{p}(0)$
and decreasing $n_{p}(0)$ [see Figs.~\ref{fig:12ws}(b) and~\ref{fig:12ws}(c)].
One can see from Tables~\ref{tab:2} and~\ref{tab:3} that the model with
a small $L$ has relatively large $\Lambda_{\rm{v}}$ and $g_{\rho}$.
Hence, the last term of Eq.~(\ref{eq:cwr}) can make a more significant
contribution in the case of small $L$, which may be the main reason for
the high $\mu_e$ and the rapid decrease of $n_p(0)$,
corresponding to small $L$ at high densities
[see Figs.~\ref{fig:13mu}(a) and \ref{fig:12ws}(c)].
Furthermore, large $\Lambda_{\rm{v}}$ and $g_{\rho}$, corresponding to
small $L$, results in high $\mu_n$ and low $\mu_p$,
as shown in Figs.~\ref{fig:13mu}(b) and \ref{fig:13mu}(c).
For the neutron density at the center, $n_{n}(0)$, and that at the boundary,
$n_{n}(r_{\rm{ws}})$, plotted in Fig.~\ref{fig:12ws}(b), it is seen that
the model with a larger $L$ predicts smaller $n_{n}(0)$ and
larger $n_{n}(r_{\rm{ws}})$, which are more pronounced at high densities.
The behaviors of $n_{n}(0)$ and $n_{n}(r_{\rm{ws}})$ obtained in the present
study are consistent with those reported in Refs.~\cite{Oyam07,Gril12}.
The $L$ dependence of $n_{n}(0)$ and $n_{n}(r_{\rm{ws}})$ can be understood
from the density dependence of the symmetry energy $E_{\rm sym}$.
In one set of generated models, $E_{\rm sym}$ has the same value at
$n_{\rm{fix}}=0.11\, \rm{fm}^{-3}$ for different $L$. However, a larger $L$
in one set of generated models corresponds to a larger $E_{\rm sym}$ at
higher density in the center region and to a smaller $E_{\rm sym}$ at lower
density in the neutron gas outside. Therefore, a larger $L$ favors a more
diffuse neutron distribution, which results in smaller $n_{n}(0)$
and larger $n_{n}(r_{\rm{ws}})$, as shown in Fig.~\ref{fig:12ws}(b).
This tendency can be seen more clearly in Fig.~\ref{fig:14wsD},
in which the density profiles are plotted with two extreme values
of $L$ in the set of TM1 at several average baryon densities.
We conclude that a larger $L$ in one set of generated models
predicts a higher neutron drip density $n_{\rm{drip}}$ due to its lower
neutron chemical potential $\mu_n$. Moreover, with increasing density,
neutrons drip out more easily for the model with a larger $L$ due to
its lower $E_{\rm sym}$ in the dilute neutron gas.
As a result, a larger value of $L$ predicts a higher neutron gas
density $n_{n}(r_{\rm{ws}})$ in the high-density region.

We show in Fig.~\ref{fig:14wsD} the density distributions of neutrons and
protons in the Wigner--Seitz cell at different average baryon density $n_b$
with two extreme values of $L$ in the set of TM1.
As $n_b$ increases, $r_{\rm{ws}}$ clearly decreases
and the neutron density outside becomes much larger.
On the other hand, the proton density at the center
decreases significantly with increasing $n_b$, while the neutron density
at the center does not change very much for different $n_b$.
However, the distributions of protons and neutrons
become more diffuse at higher density.
One can see that the differences between $L=40$ MeV and $L=110.8$ MeV
significantly increase with increasing $n_b$.
The nuclear size obtained with $L=40$ MeV is larger than that with
$L=110.8$ MeV, especially for the case of $n_b=0.05\,\rm{fm}^{-3}$,
which leads to larger $Z$ and $A_d$, as shown in Fig.~\ref{fig:11tfdrop}.
Furthermore, the neutron distributions with $L=110.8$ MeV are
more diffuse than those with $L=40$ MeV, which can be explained
by the density dependence of the symmetry energy $E_{\rm sym}$,
as discussed above. It is clearly seen that the neutron gas density
with $L=110.8$ MeV increases more rapidly than that with $L=40$ MeV,
which is also observed in Fig.~\ref{fig:12ws}(b).
Since the Coulomb interaction is self-consistently taken into account
in the TF approximation, it is seen that the proton distributions are influenced
by the Coulomb potential; namely, the proton densities at the center of the cell
are slightly lower than those at the surface region due to
the repulsive Coulomb potential.

\section{Conclusions}
\label{sec:5}

We have investigated the effects of the symmetry energy on the neutron
drip density and properties of nuclei in neutron star crusts.
The Wigner--Seitz approximation has been employed to describe the
nonuniform matter around the neutron drip density.
For the nuclear interaction, we have adopted the RMF theory
with several successful parametrizations.
We have considered and compared three different methods for
calculating properties of neutron star crusts,
namely, the self-consistent TF approximation,
the simple CP method with bulk Gibbs equilibrium conditions,
and the CLD model with equilibrium conditions determined by
including the surface and Coulomb energies.
It has been found that the simple CP method fails to describe the
nonuniform matter around the neutron drip density due to its higher
energies than that of homogeneous matter.
The failure of the CP method is mainly because the finite-size effects
due to the surface and Coulomb energies are too large to be treated
perturbatively at low densities, so that they should be included
self-consistently in determining the equilibrium state, as done in
the CLD and TF methods. The results of the CLD method have been
greatly improved by the inclusion of finite-size effects
compared to those of the CP method.
We have made a detailed comparison of the three methods
and concluded that the inclusion of surface and Coulomb
energies in determining the equilibrium state plays a
crucial role in the description of nonuniform matter
at low density.

We have examined the influence of the symmetry energy slope $L$
using two sets of models generated from the TM1 and IUFSU
parametrizations. All models in each set have the same isoscalar
saturation properties and fixed symmetry energy
at $n_{\rm{fix}}=0.11\, \rm{fm}^{-3}$, but they have different symmetry
energy slope $L$. The choice of $n_{\rm{fix}}=0.11\, \rm{fm}^{-3}$
can produce very similar binding energies for finite nuclei
within one set of generated models.
We have performed the self-consistent TF calculation to study
the influence of the symmetry energy slope $L$ on the neutron
drip density $n_{\rm{drip}}$. It has been found that $n_{\rm{drip}}$
increases with increasing $L$, which is related to the decrease of
the Wigner--Seitz cell radius. At the neutron drip point,
the proton fraction of the equilibrium nucleus is found to decrease
with increasing $L$, while the neutron skin thickness shows
an obvious increase with increasing $L$.
The $L$ dependence of the equilibrium nucleus at the neutron
drip density is qualitatively consistent with that obtained
in finite-nuclei calculations.

We have studied the effects of the symmetry energy on properties
of nuclei in the inner crust within the TF approximation.
It has been found that the proton number $Z$ and the nucleon
number $A_d$ of the droplet weakly depend on the average baryon
density $n_b$ in the low-density region, while they rapidly change
at relatively high densities. For the $L$ dependence of $Z$ and $A_d$,
it has been shown that $Z$ decreases monotonically with
increasing $L$, while $A_d$ is almost independent of $L$
at low densities. On the other hand, a strong $L$ dependence
has been observed for properties of the equilibrium nucleus
at high densities. The results obtained in the present
self-consistent TF calculation are qualitatively consistent
with those found in the literature~\cite{Oyam07,Bao14,Gril12}.
We note that nuclear shell and paring effects have been
neglected in the present work. It would be interesting
to consider these effects in future studies.

\section*{Acknowledgment}

This work was supported in part by the National Natural Science Foundation
of China (Grant No. 11375089).


\newpage
\begin{table}[tbp]
\caption{Parameter sets used in this work and corresponding
nuclear matter properties at saturation density.
The masses are given in MeV.}
\label{tab:1}
\begin{center}
\begin{tabular}{lcccc}
\hline\hline
Model          & NL3   & TM1   & FSU   & IUFSU \\
\hline
$M$            & 939.0   & 938.0   & 939.0   & 939.0   \\
$m_{\sigma}$   & 508.194 & 511.198 & 491.500 & 491.500 \\
$m_\omega$     & 782.5   & 783.0   & 782.5   & 782.5   \\
$m_\rho$       & 763.0   & 770.0   & 763.0   & 763.0   \\
$g_\sigma$     & 10.2170 & 10.0289 & 10.5924 & 9.9713  \\
$g_\omega$     & 12.8680 & 12.6139 & 14.3020 & 13.0321 \\
$g_\rho$       & 8.9480  & 9.2644  & 11.7673 & 13.5900 \\
$g_{2}$ (fm$^{-1}$) & -10.4310 & -7.2325 & -4.2771 & -8.4929 \\
$g_{3}$        & -28.885 & 0.6183  & 49.8556 & 0.4877  \\
$c_{3}$        & 0.0000  & 71.3075 & 418.3943 & 144.2195  \\
$\Lambda_{\text{v}}$ & 0.000  & 0.000  & 0.030  & 0.046 \\
\hline
$n_0$ (fm$^{-3}$) & 0.148 & 0.145 & 0.148 & 0.155 \\
$E_0$ (MeV)       & -16.3 & -16.3 & -16.3 & -16.4 \\
$K$ (MeV)         & 272   & 281   & 230   & 231 \\
$E_{\text{sym}}$ (MeV) & 37.4  & 36.9  & 32.6  & 31.3  \\
$L$ (MeV)         & 118.2 & 110.8 & 60.5  & 47.2  \\
\hline\hline
\end{tabular}%
\end{center}
\end{table}

\begin{table}[htb]
\caption{Parameters $g_\rho$ and $\Lambda_{\text{v}}$ generated from the TM1 model
for different slope $L$ at saturation density $n_0$ with fixed symmetry energy
$E_{\text{sym}}=28.05$ MeV at $n_{\rm{fix}}=0.11\, \rm{fm}^{-3}$.
The last two lines show the symmetry energy at saturation density,
$E_{\text{sym}} (n_0)$, and the neutron-skin thickness of $^{208}$Pb, $\Delta r_{np}$.
The original TM1 model has $L=110.8$ MeV.}
\label{tab:2}
\begin{center}
\begin{tabular}{lcccccccc}
\hline\hline
$L$ (MeV) & 40.0    & 50.0    & 60.0    & 70.0    & 80.0    & 90.0   & 100.0  & 110.8  \\
\hline
$g_\rho$  & 13.9714 & 12.2413 & 11.2610 & 10.6142 & 10.1484 & 9.7933 & 9.5114 & 9.2644 \\
$\Lambda_{\text{v}}$ & 0.0429 & 0.0327  & 0.0248  & 0.0182  & 0.0128 & 0.0080 & 0.0039 & 0.0000 \\
$E_{\text{sym}} (n_0)$ (MeV) & 31.38 & 32.39  & 33.29   & 34.11   & 34.86  & 35.56  & 36.22  & 36.89 \\
$\Delta r_{np}$ (fm) & 0.1574 & 0.1886  & 0.2103  & 0.2268  & 0.2402 & 0.2514 & 0.2609 & 0.2699 \\
\hline\hline
\end{tabular}%
\end{center}
\end{table}

\begin{table}[htb]
\caption{Parameters $g_\rho$ and $\Lambda_{\text{v}}$ generated from the IUFSU model
for different slope $L$ at saturation density $n_0$ with fixed symmetry energy
$E_{\text{sym}}=26.78$ MeV at $n_{\rm{fix}}=0.11\, \rm{fm}^{-3}$.
The last two lines show the symmetry energy at saturation density,
$E_{\text{sym}} (n_0)$, and the neutron-skin thickness of $^{208}$Pb, $\Delta r_{np}$.
The original IUFSU model has $L=47.2$ MeV.}
\label{tab:3}
\begin{center}
\begin{tabular}{lcccccccc}
\hline\hline
$L$ (MeV) & 47.2    & 50.0    & 60.0    & 70.0    & 80.0   & 90.0   & 100.0  & 110.0  \\
\hline
$g_\rho$  & 13.5900 & 12.8202 & 11.1893 & 10.3150 & 9.7537 & 9.3559 & 9.0558 & 8.8192 \\
$\Lambda_{\text{v}}$& 0.0460  & 0.0420  & 0.0305  & 0.0220 & 0.0153 & 0.0098 & 0.0051 & 0.0011 \\
$E_{\text{sym}} (n_0)$ (MeV) & 31.30 & 31.68  & 32.89   & 33.94  & 34.88  & 35.74  & 36.53  & 37.27  \\
$\Delta r_{np}$ (fm) & 0.1611 & 0.1739 & 0.2062  & 0.2278 & 0.2441 & 0.2571 & 0.2678 & 0.2770 \\
\hline\hline
\end{tabular}%
\end{center}
\end{table}

\newpage

\end{document}